\documentclass[3p,twocolumn,preprint]{elsarticle}

\biboptions{numbers,sort&compress}

\usepackage{graphicx}           

\usepackage{amssymb}
\usepackage{amsmath}
\usepackage{natbib}
\biboptions{numbers,sort&compress}
\usepackage{xcolor}
\usepackage{eucal}
\usepackage{hyperref}
\usepackage{soul}
\usepackage[normalem]{ulem}
\usepackage{enumitem}
\usepackage{fancyhdr}
\usepackage{hyperref}

\newcommand{\stkout}[1]{\ifmmode\text{\sout{\ensuremath{#1}}}\else\sout{#1}\fi}
\newcommand{\di}{\mathrm{d}}

\newcommand{\acc}{\mathrm{acc}}

\newcommand{\Pgen}{P_\mathrm{gen}}
\newcommand{\oPgen}{\bar P_\mathrm{gen}}

\begin{document}
\pagestyle{fancy}
\renewcommand{\headrulewidth}{0pt}
\fancyhead{} 
\fancyfoot{} 
\fancyfoot[LO,LE]{$\copyright 2022$. This manuscript version is made available under the CC-BY-NC-ND 4.0 license \href{http://www.latex-tutorial.com}{https://creativecommons.org/licenses/by-nc-nd/4.0/}}

\begin{frontmatter}

\title{Using Markov transition matrices to generate trial configurations in Markov chain Monte Carlo simulations }
 \author[1]{Jo\"el Mabillard
 \fnref{fn1}}
\author[1]{Isha Malhotra}
\author[1]{Bortolo Matteo Mognetti\corref{cor1} \fnref{fn2}}\ead{Bortolo.Matteo.Mognetti@ulb.be} 

\address[1]{Center for Nonlinear Phenomena and Complex Systems, Code Postal 231, Universit\'e Libre de Bruxelles, Boulevard du Triomphe, 1050 Brussels, Belgium}

\cortext[cor1]{Corresponding author}
\fntext[fn1]{ORCID: 0000-0001-6810-3709.} 
\fntext[fn2]{ORCID: 0000-0002-7960-8224.}

\begin{abstract} 
We propose a  new  Markov chain Monte Carlo method in which trial configurations are  generated by evolving a state, sampled from a prior distribution, using a Markov  transition matrix. We present two prototypical algorithms and derive their corresponding acceptance rules. We first identify the important factors controlling the quality of the sampling. We then apply the method to the problem of sampling polymer configurations with fixed endpoints. Applications of the proposed method range from the design of new generative models to the improvement of the portability of specific Monte Carlo algorithms, like configurational--bias schemes.

\end{abstract}

\begin{keyword}
Monte Carlo methods, Mathematical physics methods, Chemical Physics \& Physical Chemistry, Classical statistical mechanics, Markovian processes, Path sampling methods. 
\newline 
\newline
This is a post-peer-review, pre-copyedit version of an article published in Computer Physics Communications. The final authenticated version is available online at: \href{https://doi.org/10.1016/j.cpc.2022.108641}{https://doi.org/10.1016/j.cpc.2022.108641}
\end{keyword}
\end{frontmatter}
\thispagestyle{fancy}

\section{Introduction}

Markov Chain Monte Carlo (MCMC) methods are portable algorithms universally employed to sample probability functions  ($\pi$) in high--dimensional spaces~\cite{frenkel2001understanding,landau2014guide,krauth2006statistical,levin2017markov}. 
 Starting from an initial configuration, a MCMC scheme generates a sequence of states that asymptotically follow a distribution equal to $\pi$. Specifically, if $x$ is the  current configuration, a MCMC algorithm first proposes a trial state $y$ with probability $\Pgen(x\to y)$. Such a trial configuration is then accepted with probability $\acc^{(P)}(x \to  y)$, where $\acc^{(P)}$ is chosen to satisfy the detailed balance condition 
\begin{align}
J(x\to y) &= J(y\to x)
\label{Eq:DB}
\\
J(x\to y) &\equiv \pi (x) \Pgen(x \to y) \acc^{(P)}(x \to  y) \, .
\nonumber
\end{align}
The previous relations imply that $\pi$ is the stationary distribution of the Markov chain with transition matrix equal to $\Pgen(x \to y) \acc^{(P)}(x \to  y)$.

\begin{figure}[t]\centering
{\includegraphics[width=0.45\textwidth]{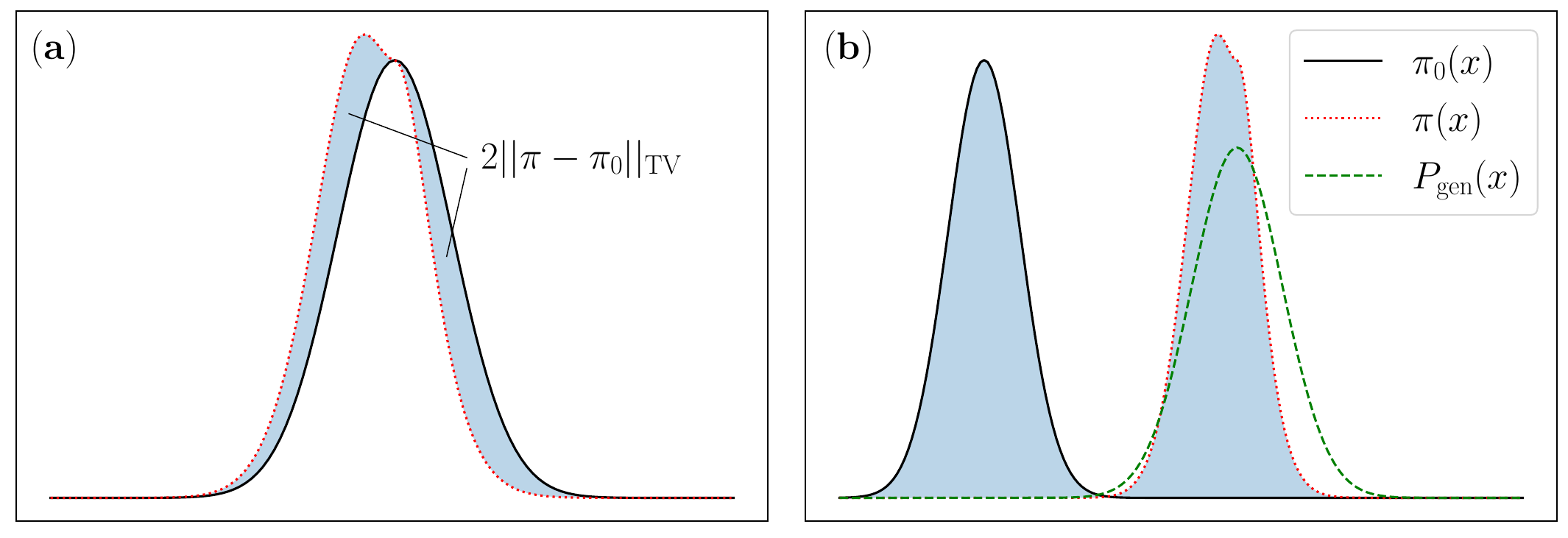}}
\caption[] {({\bf a}) In a MCMC scheme with stationary distribution $\pi$, trial configurations  can be generated by sampling a prior distribution $\pi_0$ (i.e., $P_\mathrm{gen}=\pi_0)$ if  $\pi_0$ and $\pi$ overlap (i.e., if $ \lVert \pi_0 - \pi \rVert_\mathrm{TV} = \int \mathrm{d} x |\pi(x)-\pi_0(x)|/2 \ll 1 $).  ({\bf b}) If  $\pi_0$ and $\pi$ do not overlap ($\lVert \pi_0 - \pi \rVert_\mathrm{TV}\lessapprox 1$), configurations sampled from $\pi_0$ require to be further processed before being used as trial configurations.  ({\bf a}, {\bf b}) The area of the colored regions corresponds to $2 \lVert \pi - \pi_0 \rVert_\mathrm{TV}$. 
}\label{FigIntro_new}
\end{figure}
Except for studies breaking the microscopic reversibility  condition~\cite{krauth2006statistical,bernard2009event,diaconis2000analysis,michel2020}, many developments have focused on designing  methods to generate trial configurations leading to high acceptance rates and a fast decorrelation between the configurations visited by the simulation~\cite{frenkel2001understanding,landau2014guide}. A fast decorrelation is achieved in algorithms that generate  trial configurations  that do not depend on the current state $x$, i.e., $\Pgen(x\to y)=\Pgen(y)$.  We also include within these methods the heat--bath or Glauber algorithm in which a subset of the system is updated by a direct sampling of the distribution $\pi$ conditioned on the current state of the degrees of freedom that are not touched by the move~\cite{levin2017markov}.  Trial configurations can be proposed by  direct sampling of a prior distribution $\pi_0$ if the latter overlaps with $\pi$ (Fig.~\ref{FigIntro_new}{\bf a} and Ref.~\cite{krauth2006statistical}), i.e., if $\rVert \pi - \pi_0 \rVert_\mathrm{TV} \ll 1$, where  $\rVert \cdot \rVert_\mathrm{TV}$ is the total variation distance~\cite{levin2017markov} defined as $\rVert f \rVert_\mathrm{TV}\equiv\int \mathrm{d} x |f(x)|/2$. A remarkable example is the sampling of interacting bosons represented as ring polymers~\cite{ceperley1995path}. In these systems, suitable choices of the prior $\pi_0$ allow for generating segments made of many monomers in a single update~\cite{PhysRevLett.67.2307, PhysRevB.30.2555,boninsegni2006worm}. Instead, when  $\lVert \pi - \pi_0 \rVert_\mathrm{TV} \lessapprox 1$ (see Fig.~\ref{FigIntro_new}{\bf b}),  configurations sampled from $\pi_0$ are not representative states of $\pi$. Therefore, configurations sampled from $\pi_0$ should be further processed and evolved towards $\pi$ before being used as trial configurations. In recent years, these transformations have been implemented using learned maps based on normalizing flows (e.g.,~\cite{wu2020stochastic}) resulting in generators capable of mapping smooth priors into  equilibrium states of many--body systems~\cite{noe2019boltzmann}.

\begin{figure}[t]\centering
{\includegraphics[width=0.48\textwidth]{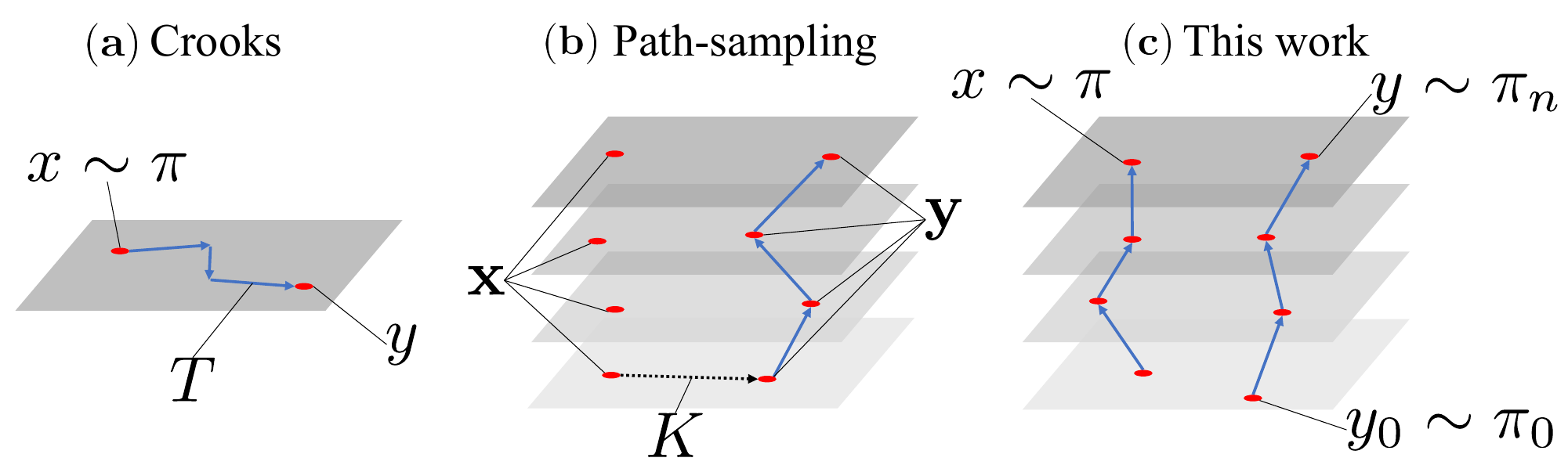}}
\caption[]{MCMC methods generating  trial configurations using Markov transition matrices ($T$): {({\bf a})} Crooks sampling~\cite{crooks1998nonequilibrium}, { ({\bf b})} a path sampling  method,~\cite{pratt1986statistical,dellago1998transition,bolhuis2002transition,ytreberg2004single,ytreberg2004erratum,athenes2004path,adjanor2005gibbs,adjanor2011waste,athenes2010free},{ ({\bf c})} the algorithms proposed in this work.
}\label{FigIntro}
\end{figure}

In this work, we introduce a class of algorithms that generate  trial configurations  that do not depend on the current state $x$, using truncated Markov Chains (tMC). Specifically, to engender a  trial configuration $y$, we first select an initial  state $y_0$ by sampling a prior distribution $\pi_0$. We then  evolve $n$ times $y_0$ using  a Markov transition matrix $T$, such that $y_i=T^i y_0$ for $i\in[1,n]$.  The distribution of the states $y_i$ is defined as $\pi_i$.  We finally identify the  trial configuration $y$ with $y_n$,  distributed as $\pi_n$,  see Fig.~\ref{FigIntro}{\bf c}.\footnote{ We do not identify $\pi_n$ with $P_\mathrm{gen}$ defined in Fig.~\ref{FigIntro_new}{\bf b} given that in the following $P_\mathrm{gen}$ will be the probability for the tMC to visit a given set of configurations $( y_0, \, y_1,\, \cdots, \, y_n)$.} The key question addressed by this study is the identification of acceptance rules satisfying Eq.~\eqref{Eq:DB} for this type of algorithm. In particular, as compared to hybrid MCMC algorithms~\cite{duane1987hybrid,mehlig1992exact,mehlig1992hybrid} based on symplectic integrators~\cite{tuckerman1992reversible,creutz1983microcanonical}, using tMCs to propose trial configurations raises difficulties as Markov transition matrices compress volumes in  configuration space. 

Notice that  tMCs have already been used in methods proposing  trial configurations by perturbing the current state $x$ (Fig.~\ref{FigIntro} {\bf a} and~\ref{FigIntro}{\bf b}). The work of Crooks~\cite{crooks1998nonequilibrium}, based on Jarzynski's results~\cite{jarzynski1997nonequilibrium,jarzynski1997equilibrium}, shows how to write acceptance rules in which $y$ is generated by  evolving $n$ times the existing configuration $x$ using $T$, see Fig.~\ref{FigIntro}{\bf a}. The results of Ref.~\cite{crooks1998nonequilibrium} cannot be generalized to the present setting in which $y$ is generated from $y_0$, see Fig.~\ref{FigIntro}{\bf c}.  Path sampling methods consider extended  configuration spaces constituted by ensembles of configurations (or paths, ${\bf x}$ and ${\bf y}$ in Fig.~\ref{FigIntro}{\bf b})~\cite{pratt1986statistical,dellago1998transition,bolhuis2002transition,ytreberg2004single,ytreberg2004erratum,athenes2004path,adjanor2005gibbs,adjanor2011waste,athenes2010free}.  Trial paths are typically proposed by first updating a single configuration (e.g., by a local transformation, $K$ in Fig.~\ref{FigIntro}{\bf b}) and then evolving it using a Markov  transition matrix $T$.  As compared to path sampling methods, the algorithms proposed in this work accept  trial configurations based on an existing configuration $x$ and not  a path. Similar to path sampling  methods, detailed balance conditions  enforce microscopic reversibility between two paths which, however, are both generated while proposing $y$.

In Sec.~\ref{Sec:Pres}, we present the proposed  MCMC method and test it using a  2D  model. 
We discuss the factors controlling the quality of the sampling by comparing  two different algorithms (A and B). Intriguingly, for Algorithm A, we show how an overlap between $\pi_n$ and $\pi$ (Fig.~\ref{FigIntro_new}{\bf b}) does not necessarily guarantee an efficient sampling. 
In Sec.~\ref{Sec:Comp}, we  further discuss similarities and differences between our scheme and path sampling methods  (Fig.~\ref{FigIntro}{\bf b}).
In Sec.~\ref{Sec:HC}, we adapt the method to the problem of sampling polymers with fixed endpoints and show how, in certain conditions, it can perform better than  a Configurational--Bias Monte Carlo (CBMC) algorithm~\cite{Siepmann1992}.  CBMC methods require performing direct sampling of a subset of degrees of freedom on the fly,  e.g., by generating polymer segments following given torsional and bending potentials. This pre--sampling task is usually addressed using ad hoc, system--dependent algorithms~\cite{vlugt1998improving,MartinSiepmann1999,Wick2000,sepehri2017improving,Sepehri2017a} while the proposed method is general and portable. Importantly, Sec.~\ref{Sec:HC} also shows how multiple tMCs can be used to propose a  single trial configuration. 
In Sec.~\ref{sec:limitations}, we  highlight the limits of the current version of the methodology and discuss some directions for improvement.
Finally, Sec.~\ref{Sec:Conc} summarises our results.

\section{Presentation of the algorithms}\label{Sec:Pres}

\subsection{Generating  trial configurations}
\label{Sec:Generate_y}

\begin{figure*}[t!]\centering
{\includegraphics[scale=0.5]{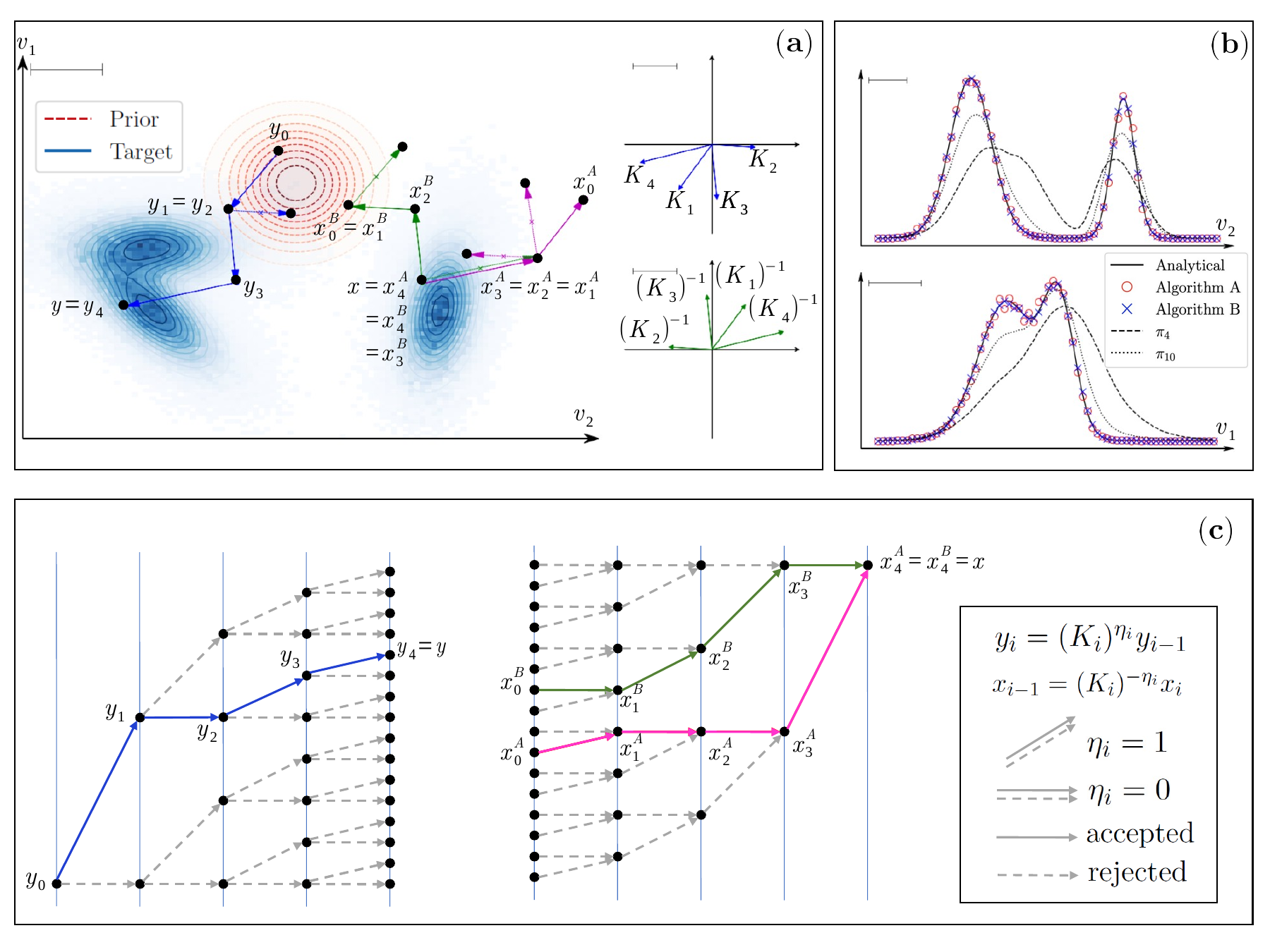}}
\caption[]{ ({\bf a}) Schematic representation of the method for a 2D model (see ~\ref{Sec:Mod2D} for details of the model and simulation parameters). Dashed and solid lines represent the level lines  of the prior, $\pi_0$, and target, $\pi$, distributions. A proposed path ${\boldsymbol{y}}_{0,n}$, is generated by accepting or rejecting a series of local displacements $K_i$ (in the panel the second displacement, $K_2$, is rejected) starting from $y_0$ distributed as $\pi_0$. Similarly, for a given state $x\equiv x_n$, the method reconstructs an extended configuration by accepting or rejecting the series of reverse transitions  $(K_n^{-1},\, K_{n-1}^{-1},\, \cdots, \, K_0^{-1})$. In the panel, Algorithm A accepts  $K_4^{-1}$ and $K_1^{-1}$, while Algorithm B accepts $K_3^{-1}$ and $K_2^{-1}$. ({\bf b}) Full lines are marginal distributions of the target distribution of panel ({\bf a}). Algorithms A and B properly sample $\pi$ (symbols). Dotted and dashed lines are the distributions of $y_n$ ($\pi_n$) with $n=10$ and  $n=4$. 
({\bf c}) Tree  representation of all possible paths starting from $y_0$ (left) and terminating in $x_n$ (right) for a given set of transformation ${\boldsymbol{K}}$. Highlighted using the color code of panel ({\bf a}) are the paths generated (left) and reconstructed  (right) by Algorithms A and B. The distance and ordering between points belonging to the same level (vertical line) have no physical meaning and multiple points may in fact correspond to the same physical configuration. In panels ({\bf a}) and ({\bf c}),  we set $n$ to $n=4$. In panels ({\bf a}) and ({\bf b}), the scale bar represents the unit length. The results  of panel ({\bf b}) have been obtained with $2\cdot10^6$ iterations.
}\label{Fig1}
\end{figure*}

\begin{figure}[t]\centering
{\includegraphics[width=0.48\textwidth]{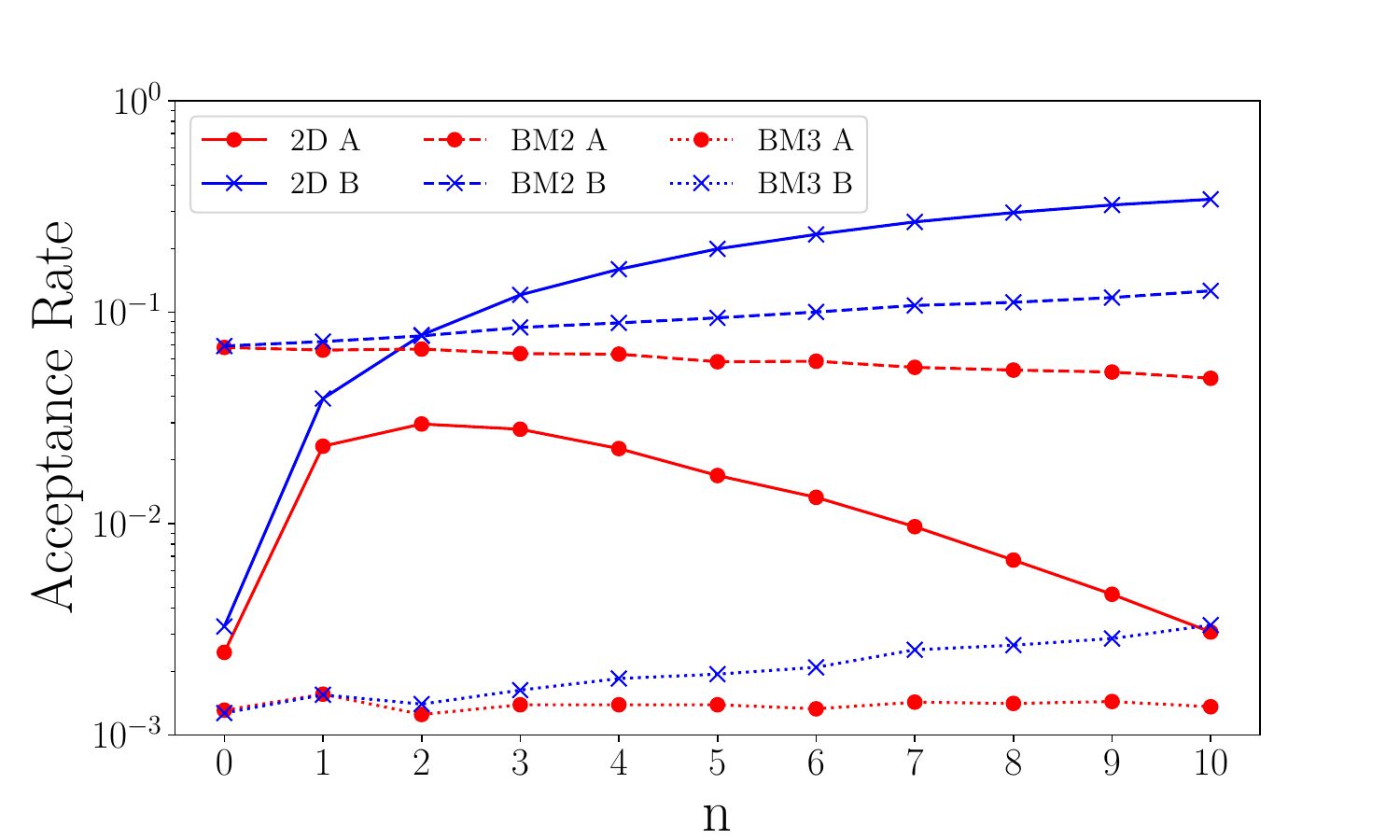}}
\caption[]{Acceptance rates of  Algorithms A and  B as  a function of the number of  states $n$ generated by the truncated Markov  chain.  We consider the 2D model of Fig.~\ref{Fig1} (2D) and the molecules with 3 and 4 branches (BM2 and BM3, see Fig.~\ref{Fig4}).
}\label{Fig2}
\end{figure}

Given a prior distribution $\pi_0$, which we assume simple enough to be sampled  statically, we  generate a starting configuration $y_0$. 
We then recursively  evolve $y_0$ using a Markov  chain with transition matrix $T$ and define $y_i=T^i y_0$ with $i\in[1,n]$.  Specifically, for a given $y_{i-1}$,  we first  propose a  trial configuration, $K_i y_{i-1}$,  using a transformation (e.g., a local translation) $K_i$ and then  accept it with  probability equal to $\mathrm{acc}^{(T)}(y_{i-1}\to K_i y_{i-1})$  such that
\begin{align}
\pi_T(y_{i-1}) T(y_{i-1} \to y_i) &=
\pi_T(y_i) T(y_i \to y_{i-1}),
\label{Eq:DB2}
\end{align}
where  $T=K_i\cdot acc^{(T)}$ and $\pi_T$ is the stationary distribution  of $T$.  This generative method is illustrated in Fig.~\ref{Fig1}{\bf a} and~\ref{Fig1}{\bf c}, left,  using a  2D model (detailed in Sec.~\ref{Sec:2DM}).

The functions $\pi$ in Eq.~\eqref{Eq:DB} and  $\pi_T$ in Eq.~\eqref{Eq:DB2} are  independent.  With the exception of the cases presented in~\ref{app:AlgosABprime}, in the remaining of the paper we  choose $\pi_T=\pi$. The transformation $K_i$ conserves volumes in  configuration space and is sampled from a generic distribution $\mu(K_i)$.   No restrictions are put on $\mu(K)$, to the extent that the ergodicity of  the MCMC method is guaranteed. In particular, the condition $\mu(K)=\mu(K^{-1})$ is not required, see Sec.~\ref{Sec:Comp}.

 We define by $\boldsymbol{y}_{0,n}$ the ensemble of configurations (in the following also labelled with path or trajectory) visited by the tMC, ${\boldsymbol{y}_{0,n}} \equiv (y_0,\, y_1,\, \cdots, y_n)$. Given the $n$ proposed transformations, ${\boldsymbol{K}}=(K_1,\, K_2, \cdots ,\, K_n)$ and the starting configuration $y_0$, the state $y_i$ is calculated as follows 
\begin{align}
y_i = (K_i)^{\eta_i}\dots (K_1)^{\eta_1} y_0 
\label{eq:xnew}
\end{align}
where $\eta_j=0$ or 1, respectively, if the transformation $K_j$ is rejected or accepted, according to $\acc^{(T)}(y_{j-1}\to K_j y_{j-1})$. There are $2^n$ (possibly degenerate) final configurations $y_n$ identified by the set of acceptances $\boldsymbol{\eta} \equiv ( \eta_1,\, \eta_2,\, \cdots,\, \eta_n )$,   see Fig.~\ref{Fig1}{\bf c}, left.  The probability of generating  ${\bf y}_{0,n}$ for a given ${\boldsymbol{K}}$ is then 
\begin{align}
\Pgen( \boldsymbol{y}_{0,n} | \boldsymbol{K} ) = \pi_0(y_0) \prod_{i=1}^n  f_{K_i}(y_{i-1},\eta_i)
\label{Eq:Pgen}
\end{align}
with 
\begin{align} 
& f_{K_i}(y_{i-1},\eta_i)=\notag\\
&\left\{\begin{array}{lll} \acc^{(T)}(y_{i-1}\to  K_{i}y_{i-1}) & \text{if} & \eta_i=1 \\ 1-\acc^{(T)}(y_{i-1}\to  K_{i}y_{i-1})& \text{if} & \eta_i=0 \end{array}\right. \label{Eq:f}
\end{align}
where $y_{i-1}$ is calculated using Eq.~\eqref{eq:xnew}. In practice, $\Pgen( \boldsymbol{y}_{0,n} | \boldsymbol{K} )$  is computed directly while generating $\boldsymbol{y}_{0,n}$, as described in ~\ref{Sec:App_Generate_y}.

Having generated the  trial configuration, it remains to identify the  acceptance rule. The probability $\Pgen( \boldsymbol{y}_{0,n} | \boldsymbol{K})$ cannot be directly identified with $\Pgen(x\to y)$ appearing in Eq.~\eqref{Eq:DB} as the latter  includes the contributions of all trajectories terminating in $y$, obtained by all possible choices of ${\boldsymbol{K}}$.  In Sec.~\ref{Sec:AlgoA} and~\ref{Sec:AlgoB}, we reconstruct trajectories ending in $x$  with the set of transformation $\boldsymbol{K}$ used to generate $y_n$. Similar to  path sampling methods, we  then assign a statistical weight to each trajectory $\boldsymbol{y}_{0,n}$ defined on an extended configuration space. Probability distributions defined over the extended  configuration space, along with $\Pgen( \boldsymbol{y}_{0,n} | \boldsymbol{K})$, allow writing  detailed balance conditions between trajectories, and, therefore, acceptance rules for $y$.

Finally, let us stress again that  trial trajectories  are not correlated with the current configuration $x$. This property is pivotal in applications as the one studied in Sec.~\ref{Sec:HC} and allows estimating the free energy of the system. Moreover, in Sec.~\ref{Sec:Comp}, we sketch out an alternative method in which  trial paths are constructed starting from the existing one. The latter method allows studying systems in which $\pi_0$ cannot be sampled  statically. 

\subsection{Algorithm A}\label{Sec:AlgoA}
We introduce the  partition function $Z^{(A)}$ defined over the ensemble of all possible trajectories ${\boldsymbol{x}}_{0,n}=(x_0,\,  x_1,\, \cdots, \, x_n)$ 
\begin{align}
	Z^{(A)}&=\int \di x_0
\left[ \prod_{i=1}^n \sum_{\eta^A_i\in \{ 0,1 \} } \di \mu (K_i)  \right] e^{-\beta H(x_n)}
\end{align}  
where  $\beta=1/(k_B\textrm{T})$, $k_B$ is the Boltzmann constant, $\textrm{T}$ the temperature, and $H$ is the target Hamiltonian $\pi(x) \sim \exp[ -\beta H(x)]$. Given the set of transformations $\boldsymbol{K}$, the configuration $x_i$, with $i\in[1,n]$, is uniquely determined by $x_0$ and  $\boldsymbol{\eta}^A=(\eta^A_1, \cdots, \eta^A_n )$ as
\begin{align}
x_i &=  (K_i)^{\eta^A_i}\dots(K_1)^{\eta^A_1} x_0\, ,
\label{eq:xold}
\end{align}
see Fig.~\ref{Fig1}{\bf c}, right.

Treating for convenience the physical variable $x_n$ as an independent variable, we  identify a state ${\boldsymbol{x}}_{0,n}$ in the extended space  by $x_n$ and  $\boldsymbol{\eta}^A$, and calculate $x_0$ by inverting Eq.~\eqref{eq:xold}.   Since $K$ conserves volumes in  configuration space, the Jacobian of the change of variables $\{x_0,\, \boldsymbol{\eta}^A\} \to \{x_n,\, \boldsymbol{\eta}^A\}$ is equal to 1, leading to the following expression for $Z^{(A)}$
\begin{align}
	Z^{(A)}&=\int \di x_n
\left[ \prod_{i=1}^n \sum_{\eta^A_i\in \{0,1 \} }\di \mu (K_i)  \right] e^{-\beta H(x_n)}\;.
\label{Eq:ZA}
\end{align}
The marginal distribution of the physical variable $x_n$ is $\pi$ and, therefore, sampling trajectories according to $Z^{(A)}$ provides configurations $x_n$ distributed as $\pi(x_n)$. For a given $x_n$ and $\boldsymbol{K}$, Eq.~\eqref{Eq:ZA} shows how all $\eta^A_i$'s are uniformly distributed with $\mathrm{Prob}(\eta^A_i=0)=\mathrm{Prob}(\eta^A_i=1)=1/2$. We stress that in Eq.~\eqref{eq:xnew}, the $\eta_i$'s are determined with an acceptance test whereas in Eq.~\eqref{eq:xold} the $\eta^A_i$'s are sampled with a uniform distribution.

 A trial configuration $y$ is generated using a set  of transformations $\boldsymbol{K}$  that does not coincide with the set of transformations used to generate the current configuration $x$. A  trajectory  $\boldsymbol{x}_{0,n}$ ending in the current configuration $x$ must then be reconstructed using the set  $\boldsymbol{K}$. Therefore, in contrast with typical path sampling  methods, for each configuration $x$ visited, the current methodology will consider two or more paths. The final state $x_n$ of the corresponding path ${\boldsymbol{x}}_{0,n}$ being identified with $x$, we sample $n$ uniformly distributed $\eta^A_i$, according to Eq.~\eqref{Eq:ZA}. The configurations $x_{i}$, for $i=n-1$ to $0$ are obtained by inverting Eq.~\eqref{eq:xold}, using iteratively $x_{i}=(K_{i+1})^{-\eta^A_{i+1}}x_{i+1}$ such that
\begin{align}
x_i=\left(\prod_{j=i+1}^n K_{j}^{-\eta^A_{j}}\right) x_n \, .\label{Eq:x_i}
\end{align} 
The process is illustrated in Fig.~\ref{Fig1}{\bf a} and~\ref{Fig1}{\bf c},  right.  Having reconstructed the trajectory ending in $x$, the probability $\Pgen(\boldsymbol{x}_{0,n}| \boldsymbol{K})$  of generating $\boldsymbol{x}_{0,n}$ using the tMC is computed as in Eq.~\eqref{Eq:Pgen}.

The  trial configuration is finally accepted with probability $\acc^{(P)}=F(z_A)$, e.g.,  $F(z_A)=\mathrm{min}(1,z_A)$ or $F(z_A)=z_A/(1+z_A)$ when using, respectively, the Metropolis or the heat--bath acceptance~\cite{frenkel2001understanding,landau2014guide} with 
\begin{align}
z_A=\frac{ \Pgen(\boldsymbol{x}_{0,n} | \boldsymbol{K})}{ \Pgen(\boldsymbol{y}_{0,n} |\boldsymbol{K}) } \frac{ \pi(y_n)}{ \pi(x_n) } \, .
\label{Eq:zA}
\end{align}
A  chart flow of Algorithm A is given in ~\ref{Sec:Pgen}.

\subsection{2D model} \label{Sec:2DM}
 To test Algorithm A, we consider  a system in which $\pi_0$ and $\pi$ do not overlap, see Fig.~\ref{Fig1}{\bf a}. $\pi_0$ is a Gaussian distribution while $\pi$ a multimodal distribution (the analytic expressions of $\pi_0$ and $\pi$ are reported in ~\ref{Sec:Mod2D}).  Existing algorithms combining local moves with (eventually learned) global maps proposing jumps between different energy minima~\cite{sbailo2021neural} would certainly outperform the presented method. The  purpose of this section is to verify that the proposed algorithms are not biased.  In particular, we have not optimised the simulation parameters (prior distribution and trial displacements).

 $K_i$ attempts to displace  a 2D vector within a square with size equal to $4$.  We consider a transition matrix $T$ that asymptotically samples $\pi$  (i.e., $\pi=\pi_T$). A case where $\pi$ is different from $\pi_T$ is studied in ~\ref{Sec:Ap}. The sampling of the target distribution obtained with  Algorithm A  (with $n=4$ and $n=10$) using $2\cdot10^6$ MC iterations~\cite{Code} is shown  on Fig.~\ref{Fig1}{\bf b} (red circles). A comparison with the analytical prediction (solid line) shows that  Algorithm A properly samples the target distribution  $\pi$.  Since $\pi=\pi_T$, the  distributions of $y_n$ ($\pi_n$, dashed and dotted lines in Fig.~\ref{Fig1}{\bf b}) and $\pi$ overlap at large values of $n$. In general, $y_n$ remains far from $\pi$  until reaching the mixing time $n_\mathrm{mix}$  (which we define as $ \rVert \sum_x \pi_0(x)  T^{n_\mathrm{mix}}(x \to \cdot) -\pi(\cdot)  \rVert_\mathrm{TV}  = 1/4$). For  $n \ll n_\mathrm{mix}$ the method cannot sample the target distribution  and features low acceptance rates (see Sec.~\ref{sec:limitations}).

\begin{figure}[t]\centering
{\includegraphics[width=0.48\textwidth]{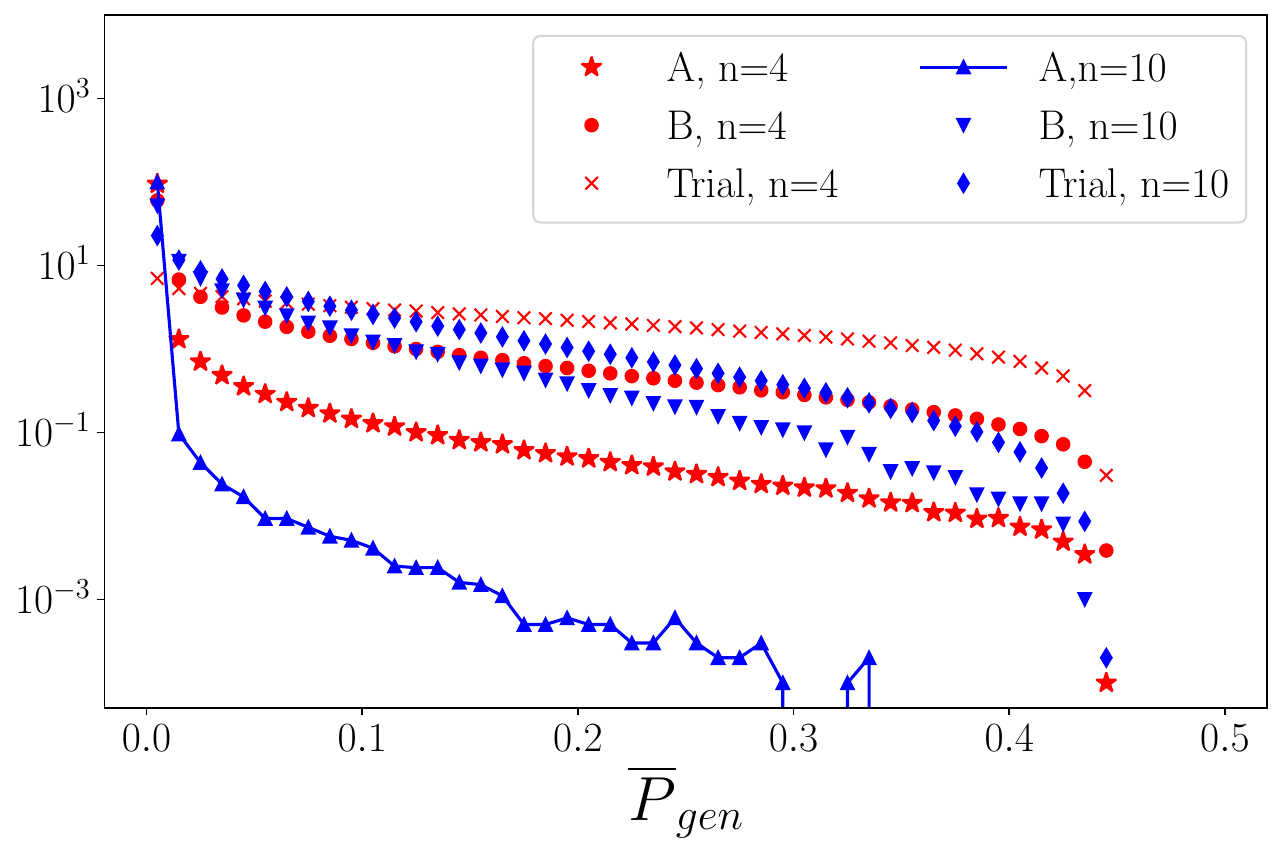}}
\caption{ Distributions of the probability of generating trial (``Trial'', see Eq.~\eqref{Eq:oPgenNew}) and  equilibrium configurations using Algorithms A and B  (``A'' and ``B'', see Eq.~\eqref{Eq:oPgenA}) for  two values of $n$. In Algorithm A,  $\boldsymbol{\eta}^A$ is uniformly distributed while in Algorithm B, $\boldsymbol{\eta}^B$ follows Eq.~\eqref{Eq:Bayes} (see ~\ref{Sec:Pn} for more details). The distributions have been calculated while producing the results of Fig.~\ref{Fig1}{\bf b}.  
}\label{Fig3}
\end{figure}

Despite the fact that  $y_n$ samples $\pi$ for $n\to \infty$ (if $\pi_T=\pi$), arbitrarily  large values of $n$  ($n>n_\mathrm{mix}$) do not improve the quality of the sampling. In particular, the acceptance of Algorithm A is non--monotonous in $n$, see  Fig.~\ref{Fig2} ``2D  A".  The poor performance of Algorithm A at large values of $n$ is explained by the  fact that $\boldsymbol{x}_{0,n}$ are random walks that are not distributed as  $\Pgen(\boldsymbol{x}_{0,n} | \boldsymbol{K})$.  For instance, in  Fig.~\ref{Fig1}{\bf a}, $x_0^A$ is found in the tail of $\pi_0$ resulting in a small value of $\Pgen(\boldsymbol{x}_{0,n} | \boldsymbol{K})$. Recall that $\boldsymbol{x}_{0,n}$ and  $\boldsymbol{y}_{0,n}$ are distributed differently: $\boldsymbol{y}_{0,n}$ is determined  by $\pi_0$ and $T$ as in Eq.~\eqref{eq:xnew}, while $\boldsymbol{x}_{0,n}$ by the extended partition functions $Z^{(A)}$ in  Eq.~\eqref{Eq:ZA}. In general, small values of $\Pgen(\boldsymbol{x}_{0,n} | \boldsymbol{K})$ are expected  for $n\to\infty$, leading to small $z_A$ in Eq.~\eqref{Eq:zA} and small acceptances. This analysis is supported by  the comparison of the distributions of the averaged probabilities of generating  trial and  equilibrium configurations, see Fig.~\ref{Fig3}. For the  trial configurations (Fig.~\ref{Fig3} ``Trial'') we have 
\begin{align}
\oPgen &=& \int \di \mu(K) \di y_0\ \pi_0(y_0) \Pgen(\boldsymbol{y}_{0,n} (\boldsymbol{\eta}) | \boldsymbol{K})
\label{Eq:oPgenNew}
\end{align}
with the  acceptances $\boldsymbol{\eta}$ calculated as in Eq.~\eqref{Eq:Pgen}. For the  equilibrium configurations we have (Fig.~\ref{Fig3} ``A'')
\begin{align}
\oPgen &=& \int \di \mu(K) \di x_n\ \pi(x_n) \Pgen(\boldsymbol{x}_{0,n}  (\boldsymbol{\eta}^A) | \boldsymbol{K})
\label{Eq:oPgenA}
\end{align}
with $\boldsymbol{\eta}^A$  sampled as described above. Comparing the case $n=4$  with $n=10$, we observe that $ \Pgen(\boldsymbol{x}_{0,n} | \boldsymbol{K})$ has on average much smaller values for larger $n$. 


\subsection{Algorithm B}\label{Sec:AlgoB}
To alleviate the problem of low values of $\Pgen(\boldsymbol{x}_{0,n} | \boldsymbol{K})$ as compared to $\Pgen(\boldsymbol{y}_{0,n} | \boldsymbol{K})$, we modify the extended partition function in Eq.~\eqref{Eq:ZA} to increase  the overlap  between the  distributions of  $\boldsymbol{x}_{0,n}$ and  $\boldsymbol{y}_{0,n}$,  and, therefore, reduce the gap between $\Pgen(\boldsymbol{x}_{0,n} | \boldsymbol{K})$ and $\Pgen(\boldsymbol{y}_{0,n} | \boldsymbol{K})$. We define the  partition function in the extended space as
\begin{align}
Z^{(B)}&=\int \di x_n  \prod_{i=1}^n  \sum_{\eta^B_i\in \{0,1 \} } \left[ \di \mu (K_i)  f_{K_i}(x_{i-1},\eta^B_i) \right] 
\nonumber \\
&\times
 \pi_0(x_0)J(x_n|\boldsymbol{K})\exp[-\beta H (x_n)],
\label{Eq:ZB}
\end{align}
 where each trajectory $\boldsymbol{x}_{0,n}$ ending in $x_n=x$ is weighted by its generating probability $\Pgen(\boldsymbol{x}_{0,n} |\boldsymbol{K})$  (see Eq.~\ref{Eq:Pgen}). The term  $J$ is a bias that constrains the marginal distribution of the physical variable $x_n$ to be equal to  $\pi(x_n)\sim \exp[-\beta H(x_n)]$. Given that $\pi_n\neq \pi$, the distributions of  $\boldsymbol{x}_{0,n}$ and $\boldsymbol{y}_{0,n}$ are not  identical, $y_n$ being  distributed as $\pi_n$ while $x_n$ as $\pi$. The bias term $J$ reads as follows
\begin{align}
J(x_n|\boldsymbol{K}) &= \left[ \sum_{\boldsymbol{\eta}^B\in \{0,1\}^n} \Pgen(\boldsymbol{z}_{0,n} | \boldsymbol{K})\delta_{z_n,x_n} \right]^{-1}
\label{Eq:J}
\end{align}
where, for ${\boldsymbol{K}}$ given,  a sum is performed over all the $2^n$ paths ending in  $x_n$, and identified by  a set of acceptances $\boldsymbol{\eta}^B$, see Fig.~\ref{Fig1}{\bf c}. In particular, the trajectory $\boldsymbol{z}_{0,n}=(z_0,\cdots,z_n=x_n)$ is constructed  by inverting Eq.~\eqref{eq:xnew} such that $z_{i-1}=K_i^{-\eta^B_i}z_i$. 

As done in Algorithm A, we reconstruct the  path $\boldsymbol{x}_{0,n}$ in the extended space by sampling the partition  function, Eq.~\eqref{Eq:ZB}. Given the set of displacements $\boldsymbol{K}$, we consider the tree made of the $2^n$ possible trajectories leading to $x_n$ as shown in Fig.~\ref{Fig1}{\bf c}, right.   A trajectory $\boldsymbol{x}_{0,n}$ is selected by sampling $\Pgen(\boldsymbol{x}_{0,n}|\boldsymbol{K})$ using Bayes' theorem. Given $x_i$, we choose $x_{i-1}$ (and,  therefore, $\eta_i^B$) among $x_i$ and  $K_i^{-1} x_i$ with probability  $P(K_i^{-\eta^B_i} x_i | x_i)$
\begin{align}
P(K_i^{-\eta^B_i} x_i | x_i) &=\frac{  f_{K_i}( K_i^{-\eta^B_i} x_i,\eta^B_i) P_{i-1}(K_i^{-\eta^B_i} x_i |\boldsymbol{K})}{  P_i(x_i|\boldsymbol{K})} 
\label{Eq:Bayes}
\end{align} 
with  $\eta^B_i=0$ or $1$ and where $P_i(x_i|\boldsymbol{K})$ is calculated recursively 
\begin{align}
P_i(x_i|\boldsymbol{K}) &=   f_{K_i}( x_i,0) P_{i-1}(x_i|\boldsymbol{K})
\nonumber \\
& +  f_{K_i}( K_i^{-1} x_i,1) P_{i-1}(K_i^{-1} x_i|\boldsymbol{K})\label{eq:Pn}
\end{align}
with $ f_{K_i}$ defined in Eq.~\eqref{Eq:f}.  $P_{i}(x_i|\boldsymbol{K})$ is the probability to visit the state $x_i$ when sampling layer $i$ at a given $(\eta^B_{i+1},\,\eta^B_{i+2},\cdots, \, \eta^B_n)$, $\boldsymbol{K}$ and $x_n$.\footnote{Notice that some within the $2^{n-i}$ possible states at layer $i$ may coincide. That is often the case when considering discrete systems. In that case, $P_i(y_i)$ is the probability of sampling $y_i$ divided by the multiplicity of $y_i$ at a given  $\boldsymbol{K}$ and $x_n$.} In particular, we have $P_{0}(x_0|\boldsymbol{K})=\pi_0(x_0)$ and  $P_n(x_n|\boldsymbol{K})=J(x_n|\boldsymbol{K})^{-1}$. Importantly, the calculation of  $P_n$ limits the algorithm to small values of $n$ given the necessity of enumerating $2^n$ states. An algorithm with polynomial computational complexity will be presented elsewhere.

After sampling  the $\eta_i^B$'s with the iterative use of Eq.~\eqref{Eq:Bayes}, the probability of selecting $\boldsymbol{x}_{0,n}$ is still given by Eq.~\eqref{Eq:Pgen}. The  trial configuration is then accepted with probability $\acc^{(P)}=F(z_B)$ with 
\begin{align}
z_B= \frac{J(y_n|\boldsymbol{K}) \pi(y_n)}{ J(x_n|\boldsymbol{K}) \pi(x_n) } = \frac{P_n(x_n|\boldsymbol{K}) \pi(y_n)}{ P_n(y_n|\boldsymbol{K}) \pi(x_n) }\, .
\label{Eq:zB}
\end{align}
 Interestingly, $z_B$ is not a function of $\Pgen(\boldsymbol{x}_{0,n}|\boldsymbol{K})$ and $\Pgen(\boldsymbol{y}_{0,n}|\boldsymbol{K})$ as they cancel the corresponding terms appearing  in the distribution of the extended  configurations, see Eq.~\eqref{Eq:ZB}.

In Fig.~\ref{Fig1}{\bf b} (blue crosses), we verify that Algorithm B is not biased in reproducing the target distribution (solid black line) for the  2D model of Sec.~\ref{Sec:2DM}. The acceptance rate now increases with $n$, as seen on Fig.~\ref{Fig2}, ``2D B'' (blue solid line and crosses). This improved acceptance rate is due to higher values of $\oPgen$,  as can be seen  by comparing  ``A'' with ``B'' on Fig.~\ref{Fig3}. In this Figure, $\oPgen$ for Algorithm B is calculated using Eq.~\eqref{Eq:oPgenA} with the  acceptances $\eta^B_i$'s obtained from Eq.~\eqref{Eq:Bayes}.   The chart flow of Algorithm B with details about the computation of $P_n$ are given in~\ref{Sec:Pn}.

We  note that the calculation of $P_n(y_n| \boldsymbol{K})$ allows sampling the excess free energy $\Delta F$ of the targeted system, defined as $\Delta F\equiv F_T- F_0$,
\begin{eqnarray}
e^{-\beta \Delta F} = {\int \mathrm{d} x \, \exp [- \beta H(x)] \over \int \mathrm{d} x \, \exp [- \beta H_0(x)] } = {Z^{(B)} \over Z^{(0)}} \,.
\label{Eq:DF0}
\end{eqnarray}
If $\pi_0(x_0)\sim \exp[-\beta H_0(x_0)]$,  $Z^{(0)}$ can be written as
\begin{align} 
Z^{(0)}&=
\int \prod_{i=1}^n  \sum_{\eta_i\in \{0,1\}} \di x_n   e^{-\beta H_0(x_0)}  
\notag \\
&
\times\Big[ \di \mu (K_i)  f_{K_i}(x_{i-1},\eta_i) \Big] 
\label{Eq:Z0adv}
\end{align}
given that $\mathrm{d} x_n = \mathrm{d} x_0$,  since $K_i$ conserves volumes in the  configuration space. In Eq.~\eqref{Eq:Z0adv} we have also used the fact that  the sum of the probabilities of generating all possible paths emanating from a given $x_0$ at a given ${\boldsymbol K}$ is equal to 1, i.e.,
\begin{eqnarray}
&\prod_{i=1}^n  \sum_{\eta_i\in \{0,1\}} 
 f_{K_i}(x_{i-1},\eta_i) = 1\;. &
\nonumber
\end{eqnarray}
The previous considerations and Eq.~\eqref{Eq:ZB} allow rewriting Eq.~\eqref{Eq:DF0} as  
\begin{eqnarray}
e^{-\beta \Delta F} = \langle 
\exp[-\beta H(y_n)] J(y_n | \textbf{K})
\rangle_{{\bf y}_{0,n}} \, ,
\label{Eq:DF}
\end{eqnarray}
where the average is calculated using the ensemble of paths obtained in the generative method of  Sec.~\ref{Sec:Generate_y}.

As already observed, Algorithms A and B are not peculiar to the use of transition matrices $T$ having  $\pi$ as asymptotic state. This property is crucial in cases where the evaluation of $\pi$ is computationally expensive  and could be approximated by a less complex function~\cite{frenkel2001understanding}.  As a proof of principle, in ~\ref{app:AlgosABprime}, we describe and validate the case in which  $\pi_T$ is constant, using the 2D  model of Fig.~\ref{Fig1}. 

\begin{figure}[t]\centering
{\includegraphics[width=0.48\textwidth]{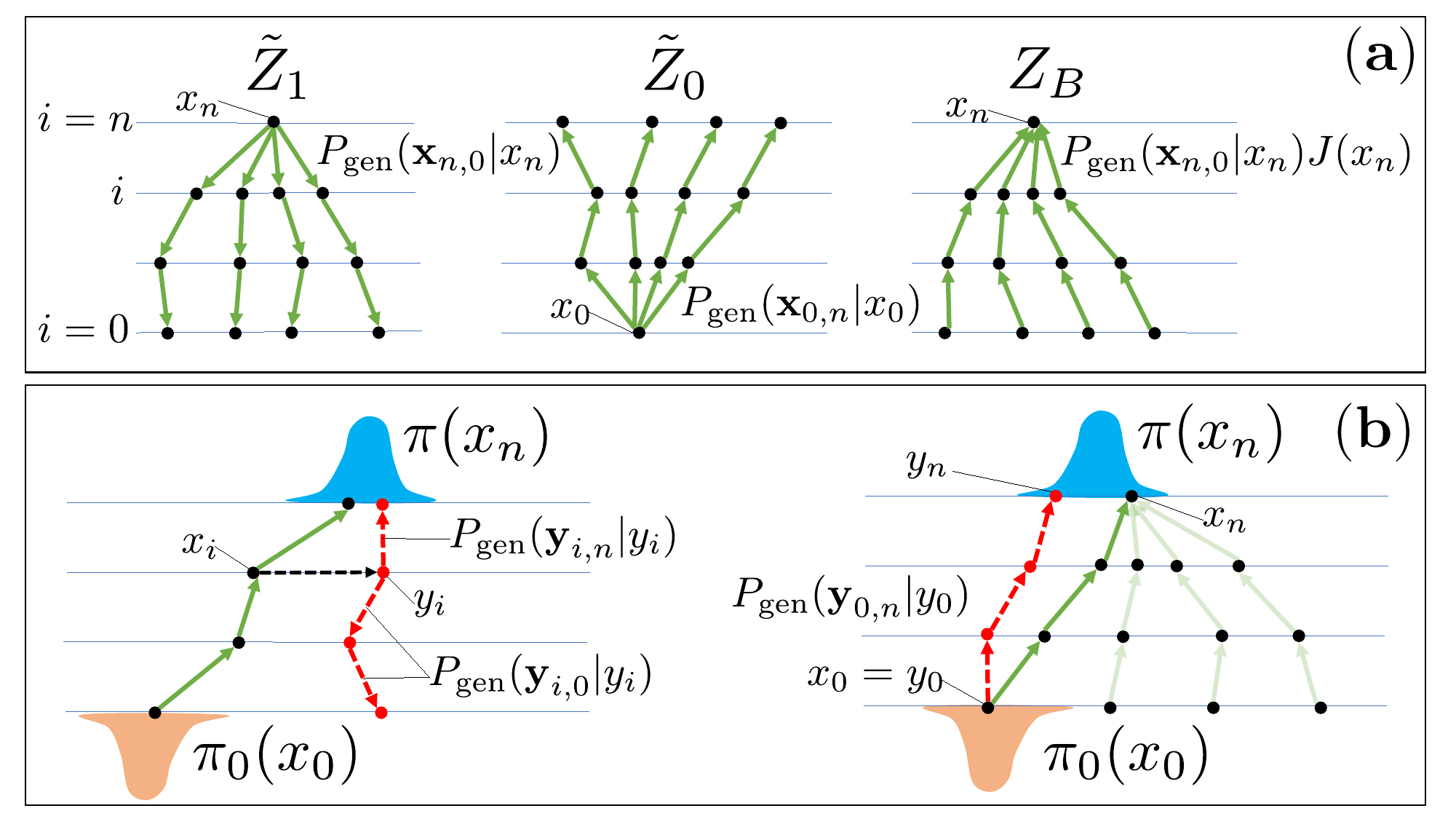}}
\caption[]{ ({\bf a})	Ensemble of paths considered in the extended partition functions presented in Ref.~\cite{athenes2004path,adjanor2005gibbs} (left and center) and in the current study (right). ${\boldsymbol{x}}_{0,n}$ and ${\boldsymbol{x}}_{n,0}$ represent trajectories generated using, respectively, the  direct and  reverse protocol~\cite{crooks1998nonequilibrium}.  
({\bf b}) (left)  Local update of a path presented in Ref.~\cite{dellago1998transition,bolhuis2002transition}. ${\boldsymbol{y}}_{i,n}$ and ${\boldsymbol{y}}_{i,0}$ are paths obtained by  using, respectively, the direct and reverse protocol. (right) Proposed method  that does not rely on  reverse protocols to generate  trial paths.   
}\label{Fig:PS}
\end{figure}
 \subsection{Comparison with other path sampling methods}\label{Sec:Comp}
Path sampling methods have been used to calculate free energies~\cite{ytreberg2004single,ytreberg2004erratum,athenes2004path,adjanor2005gibbs}. Based on Jarzynski's results~\cite{jarzynski1997equilibrium,jarzynski1997nonequilibrium}, the difference in free energy between two systems with Hamiltonian $H_0$ ($\pi_0(x)\sim \exp[-\beta H_0 (x)]$) and  $H$ ($\pi(x)\sim \exp[-\beta  H (x)]$) can be sampled using
\begin{align}
    \Delta F=-k_B T \log \langle \exp (-W) \rangle_\mathrm{path} \, ,
\label{Eq:JN}
\end{align}
where $W$ is the work performed by a protocol  (in the following labelled direct protocol) switching the Hamiltonian of the system from  $ H_0$ to $ H$ in $n$ steps, and the average is taken over all paths engendered by the protocol.

 Developments in path sampling methods focused on finding extended partition functions promoting trajectories that contribute the most to the average in  Eq.~\eqref{Eq:JN}.  For instance,  Ref.~\cite{athenes2004path,adjanor2005gibbs} considered umbrella ensembles ($\tilde Z_\theta$) interpolating the two following extended partition functions (see Fig.~\ref{Fig:PS}{\bf a}, left and center)
\begin{align}
 \tilde Z_1 &=
\int \mathrm{d} {\boldsymbol{x}} \, \pi(x_n) \Pgen ( {\boldsymbol{x}}_{n,0}|x_n) 
\label{Eq:ZT1}
\\
 \tilde Z_0 &=
\int \mathrm{d} {\boldsymbol{x}} \, \pi_0(x_0) \Pgen( {\boldsymbol{x}}_{0,n}|x_0)
\label{Eq:ZT0}
\end{align}
where $\Pgen ({\boldsymbol{x}}_{0,n}|x_0)$ is the probability of generating a given path  ${\boldsymbol{x}}_{0,n}$ starting from $x_0$, while $\Pgen ({\boldsymbol{x}}_{n,0}|x_n)$ is the probability of generating the trajectory ${\boldsymbol{x}}$ from $x_n$ by the  reverse protocol, driving the system from $H$ to $H_0$~\cite{jarzynski1997equilibrium,jarzynski1997nonequilibrium,crooks1998nonequilibrium}.

Instead, using a notation similar to Eqs.~\eqref{Eq:ZT1} and~\eqref{Eq:ZT0}, $Z_B$ introduced in Eq.~\eqref{Eq:ZB} would read as follow (see Fig.~\ref{Fig:PS}{\bf a}, right)
\begin{align}
      Z_B &=
\int \mathrm{d} {\boldsymbol{x}} \, \pi(x_n) J(x_n) \Pgen({\boldsymbol{x}}_{0,n}|x_n) \, .
\label{Eq:ZBnew}
\end{align}
Notice that 
\begin{align}
    \int \mathrm{d} x_1 \cdots \mathrm{d} x_{n-1} \Pgen ({\boldsymbol{x}}_{n,0}|x_n) &= 1
    \label{Eq:xmxn}
    \\
    \int \mathrm{d} x_2 \cdots \mathrm{d} x_{n} \Pgen ({\boldsymbol{x}}_{0,n}|x_0) &= 1
    \label{Eq:xpx0}
\end{align}
while
\begin{align}
    \int \mathrm{d} x_1 \cdots \mathrm{d} x_{n-1} \Pgen ({\boldsymbol{x}}_{0,n}|x_n) & \neq  1 \, .
    \label{Eq:xpxn}
\end{align}
The previous equations explain why in the definition of $\tilde Z_0$ in Eq.~\eqref{Eq:ZT0}, one does not need to use the term $J$  as done in Eq.~\eqref{Eq:ZBnew} to constrain the distribution of $x_n$ to $\pi$, see Sec.~\ref{Sec:AlgoB}. A similar conclusion follows for $\tilde Z_1$ with Eqs.~\eqref{Eq:ZT1} and~\eqref{Eq:xpx0}. 

As explained in Sec.~\ref{Sec:AlgoB}, the motivation for choosing an extended partition function as in Eqs.~\eqref{Eq:ZB} and~\eqref{Eq:ZBnew} is to maximize  the overlap between the distributions of trial and equilibrium trajectories (${\boldsymbol{y}}_{0,n}$ and ${\boldsymbol{x}}_{0,n}$, using the notation of Sec.~\ref{Sec:AlgoB}). In this respect, a key difference between our algorithms and existing  path sampling methods is that we never propose  trial configurations by  the reverse protocol. A typical move to propose trial configurations using  the reverse protocol is shown in Fig.~\ref{Fig:PS}{\bf b}, left~\cite{dellago1998transition,bolhuis2002transition}. In this setting,  trial trajectories are generated by a local update of one of the configurations belonging to the current path ($x_i \to y_i$), followed by propagating $y_i$ to $y_n$  and $y_i$ to $y_0$  using the direct and reverse protocol, respectively. 
 In our setting, in which we instantaneously drive the system from $H_0$ to $H$,  the reverse protocol is very inefficient in sampling $Z_B$  given that reverse and direct protocols coincide~\cite{crooks1998nonequilibrium}. It follows that the distribution of $y_0$ (obtained using the reverse protocol) would resemble more to $\pi$ than $\pi_0$,  resulting in poor sampling when $\rVert \pi-\pi_0 \rVert_\mathrm{TV}\lessapprox 1$, Fig.~\ref{Fig:PS}{\bf b}, left.
For the system of Fig.~\ref{Fig:PS},  reverse protocols will be outperformed even by  a random reconstruction of the path (as in Algorithm A, Sec.~\ref{Sec:AlgoA}) since in the latter case  $y_0$ is generated from a random walk which is not constrained by $\pi$.

As our approach does not involve  reverse protocols, the condition $\mu(\boldsymbol{K})=\mu(\boldsymbol{K}^{-1})$, which is necessary  to enforce the detailed balance condition in such a method, is not required. To support this statement,   in~\ref{app:ASdisplacement}, we sample the 2D model of Fig.~\ref{Fig1} using an asymmetric distribution of the trial displacements,  $\mu(\boldsymbol{K}) \neq \mu(\boldsymbol{K}^{-1})$.

It is worth pointing out that our method does not necessarily require the ability to sample  the prior $\pi_0$ statically. A possible algorithm generating  trial paths $\boldsymbol{y}_{0,n}$  by updating the current configuration  $x_n$ is illustrated in Fig.~\ref{Fig:PS}{\bf b}, right.  In particular, starting from  $x_n$, we  generate a set of transformations $\boldsymbol{K}$ and reconstruct the most likely path $\boldsymbol{x}_{0,n}$ (as done in Sec.~\ref{Sec:AlgoB}) using  $\boldsymbol{K}$. We then  generate a trial path $\boldsymbol{y}_{0,n}$ by a forward propagation of $x_0$ identified with $y_0$ using  $\boldsymbol{K}$. A new set of transformation $\boldsymbol{K}'$ is used in the following move. This algorithm generates more correlated configurations (as compared to the method presented in Sec.~\ref{Sec:AlgoB}) but with a higher acceptance  rate given that, in this case,  $y_0$ asymptotically follows the path distribution defined by $Z_B$ (note that in Eq.~\eqref{Eq:ZB} and~\eqref{Eq:ZBnew}, $x_0$ is not distributed as $\pi_0$).

\begin{figure*}[t]\centering
    {\includegraphics[width=1.0\textwidth]{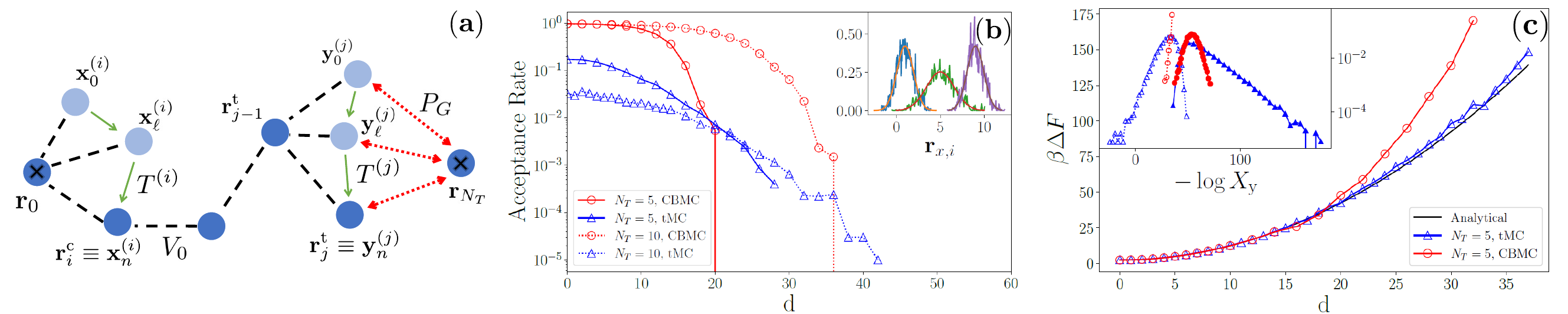}}
	\caption[]{ {\scalebox{}{}  ({\bf a}) Schematic of the truncated Markov  chain strategy employed to sample polymers with fixed endpoints. ({\bf b}) Comparison of the acceptance rates obtained by  CBMC (with $k=1000$ trials) and Algorithm B (with $n=10$) for a chain with $6$ and $11$ monomers ($N_T=5$ and $11$) as a function of the end--to--end distance $d$. Inset: Asymptotic distributions of the $i^{th}$ monomer along the  end--to--end direction ${\bf r}_{x,i}$ (with $i=1$, 5, 9, and $N_T=10$) sampled by Algorithm B compared with the expected  distributions (solid lines). ({\bf c}) $\Delta F$ calculated using Eqs.~\eqref{Eq:X_tMC} and~\eqref{Eq:X_CBMC} as compared to the analytic prediction of Eq.~\eqref{Eq:F_exp}. Inset: distribution of $X_\mathrm{y}$ (y=CBMC and y=tMC) using  current (open symbols) and  trial (full symbols) configurations  for $d=20$.
	}
}
        \label{Fig:HS}
\end{figure*}

\section{Sampling polymers with fixed endpoints}\label{Sec:HC}

We develop a method to generate chains with fixed endpoints based on Algorithm B  (Sec.~\ref{Sec:AlgoB}). This problem underlies efficient sampling of  the configurational entropy of polymers and is usually addressed using  CBMC simulations~\cite{Wick2000}. In this example, multiple tMCs are used to propose a  trial configuration. 

We consider a chain with $N_T+1$ monomers located at positions ${\bf r}_i$, $i=0, \cdots N_T$, and endpoints fixed at a distance equal to $d$,  such that $|{\bf r}_{N_T}-{\bf r}_0|=d$. Neighboring monomers interact via a harmonic potential $V_0$.  The configurational energy of the system is given by 
\begin{align}
U_{HS} &=  \sum_{i=1}^{N_T} V_0({\bf r}_i, {\bf r}_{i-1})\, .
\label{Eq:UHS}
\end{align}
 Details about the system are reported in ~\ref{Sec:ModHS}.

As in CBMC,  trial configurations are generated one monomer at a time. We use $N_T-1$ different  tMCs,  with transition matrices given by $( T^{(1)}, \cdots, T^{(N_T-1)} )$, to sequentially generate  trial monomers,  ${\bf r}^\mathrm{t}_i \equiv {\bf y}^{(i)}_n$ with $i=1,\cdots , N_T-1$. The use of multiple Markov  chains can leverage the fact that the interactions between  subsets of degrees of freedom could be sufficiently weak to be sampled perturbatively.  For instance, in the present case, only neighbouring monomers interact, see Eq.~\eqref{Eq:UHS}.

Given ${\bf r}^t_{j-1}={\bf y}^{(j-1)}_n$, we first sample ${\bf y}^{(j)}_0$  from $\pi_0({\bf y}^{(j)}_0)=\exp[- \beta V_0({\bf y}^{(j)}_0, {\bf r}^\mathrm{t}_{j-1})]$.  A truncated Markov  chain with transition matrix $T^{(j)}$, attempting to displace a monomer within a cube of size $\Delta x =4$ \footnote{$\Delta x$ should be sufficiently big to generate sufficiently stretched configurations,  $\Delta x \sqrt{n}/2 > d_\mathrm{max}/N_T$, where $d_\mathrm{max}$ is the maximal end--to--end distance considered in Fig.~\ref{Fig:HS}{\bf b}. We did not further optimize $\Delta x$.}, is then used to  evolve ${\bf y}^{(j)}_0$ towards ${\bf y}^{(j)}_n$, i.e., ${\bf y}^{(j)}_n=(T^{(j)})^n {\bf y}^{(j)}_0$ see Fig.~\ref{Fig:HS}{\bf a}. The stationary distribution of $T^{(j)}$, $\pi_{T^{(j)}}$, is taken equal to $\pi_{T^{(j)}}(y^{(j)}) \sim P_G({\bf y}^{(j)}_0,{\bf r}_{N_T}) \pi_0({\bf y}^{(j)}_0)$, where $P_G$ is a guiding function biasing the chain's growth towards the fixed end monomer~\cite{Wick2000,frenkel2001understanding}. $P_G$ is chosen as the end--to--end distance distribution of a chain segment of length $N_T-j$ with unconstrained end--to--end distance. Given the set of  the current monomers, $({\bf r}^\mathrm{c}_1,\cdots,{\bf r}^\mathrm{c}_{N_T-1} )$, we reconstruct the tree of possible trajectories for each monomer using $T^{(i)}$. Fig.~\ref{Fig:HS}{\bf a} reports one of these trajectories for monomer $i$. Each monomer contributes to the acceptance factor, $F(z_\mathrm{tMC})$, with a term given by Eq.~\eqref{Eq:zB}. In particular 
\begin{align}
z_\mathrm{tMC} &= \frac{\exp(- \beta U^\mathrm{t}_{HS})}{\exp(- \beta U^\mathrm{c}_{HS})} \prod_{i=1}^{N_T-1} \frac{P_n({\bf r}^\mathrm{c}_i|\boldsymbol{K}^{(i)})}{P_n ({\bf r}^\mathrm{t}_i|\boldsymbol{K}^{(i)})} \, ,
\label{Eq:HCztMC}
\end{align}
where $P_n({\bf r}^\alpha_i|\boldsymbol{K}^{(i)})$,  with $\alpha$ standing for $t$ or $c$, is calculated using the tree engendered by $T^{(i)}$,  as in Fig.~\ref{Fig1}{\bf c} right, while $U^\mathrm{t}_{HS}$/$U^\mathrm{c}_{HS}$ is the configurational energy of the trial/current configuration given by Eq.~\eqref{Eq:UHS}. In Eq.~\eqref{Eq:HCztMC}, $\boldsymbol{K}^{(i)}=( K^{(i)}_1,\, \cdots, \,K^{(i)}_n )$ is the set of transformations used by $T^{(i)}$.

We compare our method with a standard CBMC algorithm in which monomer ${\bf r}^\mathrm{t}_i$ is selected from $k$ trials, distributed as $\pi_0$ (${\bf r}^{(\alpha)}_i$, $\alpha=1, \cdots k$), using  $P_G$
\begin{align}
\mathrm{Prob}({\bf r}^\mathrm{t}_i)& = \frac{ P_G({\bf r}^\mathrm{t}_i, {\bf r}_{N_T}) }{ W^t_i}\;, \notag\\
 W^t_i&=\sum_{\alpha=1}^k P_G({\bf r}^{(\alpha)}_i, {\bf r}_{N_T})\;,
\end{align}
where  $W^t_i$ is the contribution of monomer $i$ to the Rosenbluth weight of the  trial configuration.  The Rosenbluth weight of the current configuration, $W^c_i$ is defined similarly~\cite{frenkel2001understanding}. The acceptance, $F(z_\mathrm{CBMC})$, reads as follows
\begin{align}
z_\mathrm{CBMC} &= \frac{e^{- \beta V_0({\bf r}^\mathrm{t}_{N_T-1},{\bf r}_{N_T})}}{ e^{- \beta V_0({\bf r}^\mathrm{c}_{N_T-1},{\bf r}_{N_T})} } \prod_{i=1}^{N_T-1}\left[ \frac{W^\mathrm{t}_i}{ W^\mathrm{c}_i } \frac{P_G({\bf r}^\mathrm{c}_i, {\bf r}_{N_T})}{P_G({\bf r}^\mathrm{t}_i, {\bf r}_{N_T})}
\right].
\nonumber 
\end{align}

The  simulation results are shown on Fig.~\ref{Fig:HS}{\bf b}. We observe how CBMC is more efficient for small values of $d$. However, when increasing $d$, the acceptance of CBMC plummets while  tMCs can more easily generate overstretched configurations. This is more evident in systems with small values of $N_T$. In the overstretched regime, CBMC fails since  $\pi_0$ and $\pi$ do not overlap. Instead,   tMCs can sample distributions  that do not overlap with $\pi_0$, as explicitly shown in Fig.~\ref{Fig1}{\bf a} and~\ref{Fig1}{\bf b}. For $d=10$, the inset of Fig.~\ref{Fig:HS}{\bf b} shows how the asymptotic distributions of ${\bf r}_i$ (with $i=1$, 5, 9, and $N_T=10$) sampled by Algorithm B follow the expected distributions.

As anticipated in Sec.~\ref{Sec:AlgoB}, 
the sampling of $P_n({\bf r}^\mathrm{t}_i|\boldsymbol{K}_i)$ (and, therefore, $J$) allows calculating the excess free energy $\Delta F$ of tethering monomer $N_T-1$ to $N_T$. The expected expression of $\Delta F$ is the following
\begin{eqnarray}
\beta \Delta F_\mathrm{exp} &=& {3\over 2} \log{N_T} + {1\over 2 N_T \sigma^2} ({\bf r}_{N_T} - {\bf r}_0)^2.
\nonumber \\
\label{Eq:F_exp}
\end{eqnarray}
By generalising the arguments leading to Eq.~\eqref{Eq:DF}, $\Delta F$ is calculated as follows when using tMCs 
\begin{eqnarray}
e^{- \beta \Delta F } &=&  \langle X_\mathrm{tMCs} \rangle_{({\bf r}^\mathrm{t}_i)_i} \;,
\\
X_\mathrm{tMCs} &=& e^{ -\beta U^t_{HS}} \prod_{i=1}^{N_T-1} {1 \over P_n({\bf r}^\mathrm{t}_i|\boldsymbol{K}_i)} \;.
\label{Eq:X_tMC}
\end{eqnarray}
Instead, the estimator used in CBMC ($X_\mathrm{CBMC}$) reads as follows~\cite{frenkel2001understanding} 
\begin{align}
X_\mathrm{CBMC} =  e^{- \beta V_0({\bf r}^\mathrm{t}_{N_T-1},{\bf r}_{N_T})} \prod_{i=1}^{N_T-1} {W^\mathrm{t}_i \over k  P_G({\bf r}^\mathrm{t}_i, {\bf r}_{N_T})}\, . 
\label{Eq:X_CBMC}
\end{align}
Fig.~\ref{Fig:HS}{\bf c} shows how the two estimators, Eqs.~\eqref{Eq:X_tMC} and~\eqref{Eq:X_CBMC}, properly reproduce $\Delta F_\mathrm{exp}$ at small values of $d$. Discrepancies appear concomitantly with the downfall of the acceptance rates.  This is expected and can be quantified by a poor  overlap between the  trial and current distributions of $X$ (see inset of Fig.~\ref{Fig:HS}{\bf c}): $\Delta F$ cannot be reproduced when the generative method cannot produce representative configurations of $\pi(x)$.

Notice that the computational complexity of the CBMC with $k=1000$ is comparable  with that of Algorithm B  with $n=10$, as employed in Fig.~\ref{Fig:HS}{\bf b}. We stress that a more efficient CBMC algorithm would generate trial segments distributed as  $c \pi \pi_0$, where $c$ is a normalization constant. However, sampling $\pi \pi_0$ would require developing system--specific sampling procedures~\cite{Wick2000}. On the other hand, the  tMC method can readily be employed for any type of potentials (including bending and torsional terms). In that sense,  tMC algorithms are more portable.

\begin{figure}[t]\centering
    {\includegraphics[width=0.48\textwidth]{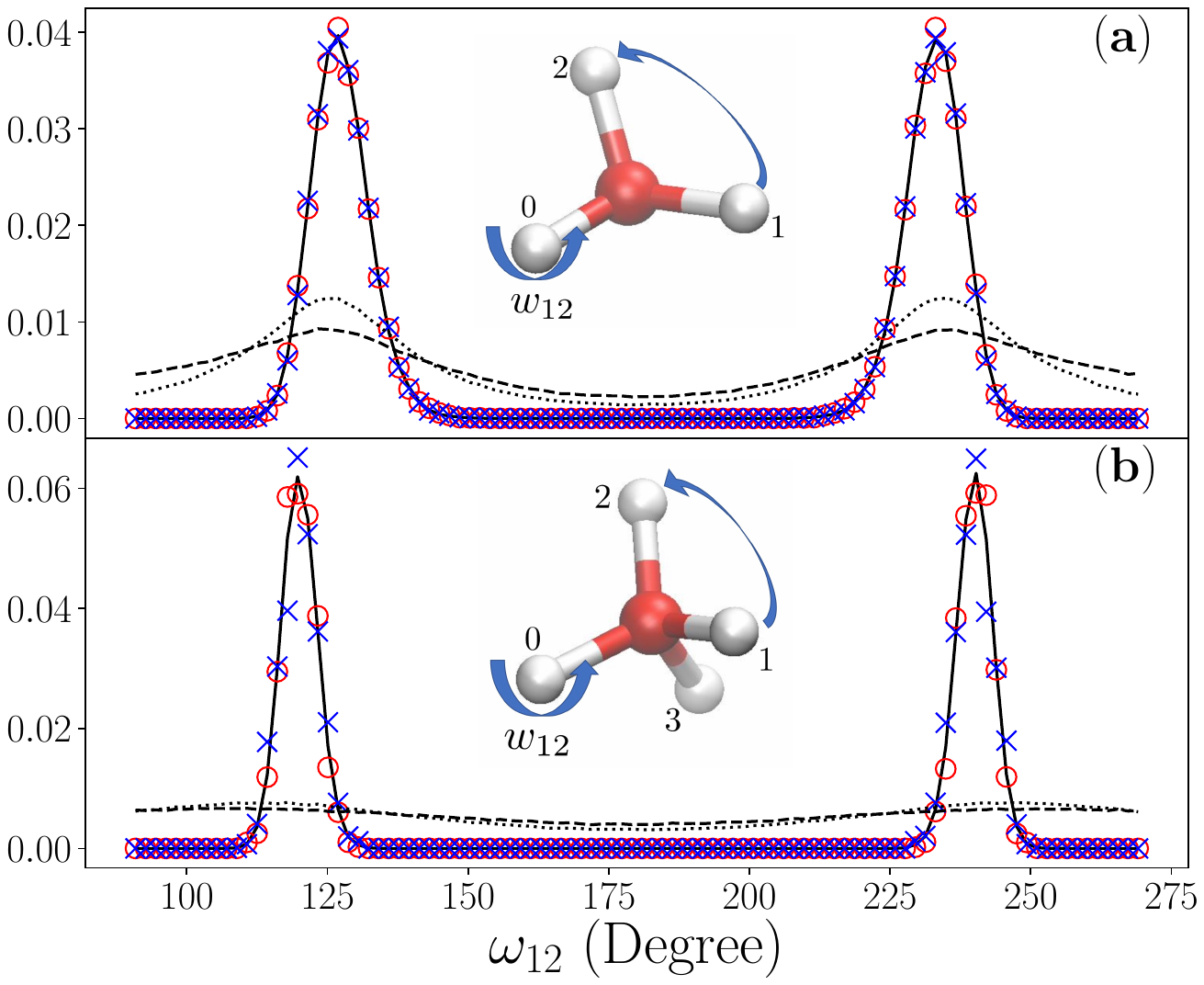}}
	\caption[]{ Simulated (symbols) and expected distribution (full line)  of  the  dihedral angle ($w_{12}$) of molecules with 3 branches, 2--methylpropane, panel {\bf a}) and 4 branches (2,2--dimethylpropane, panel {\bf b}. Dotted and dashed lines are the distributions of $y_n$ with $n=10$ and $n=4$, respectively.  The legend is as in Fig.~\ref{Fig1}{\bf c}.  The results are obtained with $10^6$ iterations.  More details of the models and simulation parameters are reported in  ~\ref{Sec:ModMol} and in~\cite{Code}.
}
        \label{Fig4}
\end{figure}


\section{Limitations of the current algorithms}\label{sec:limitations}

We sample $2$--methylpropane (BM2) and  $2,2$--dimethylpropane (BM3) molecules  modeled, respectively,  as a 3-- and 4--branched molecules (see Fig.~\ref{Fig4}). The length of the branches is fixed while  pairs of branches interact through  a  bending potential $U_\mathrm{bend}$. We fix the  the direction of a branch, and label the branches that are sampled from $1$ to $n_b$, where $n_b=2$ for BM2 or $3$ for BM3. The target  probability density function is given as
\begin{align}
\pi &= c \prod_{i<j} \exp[-\beta U_\mathrm{bend}(\theta_{ij})] \, ,
\end{align}
where  $\theta_{ij}$ is the angle between the branches $i$ and $j$, and $c$ a normalization constant.
We choose the following prior  probability density function 
\begin{align}
\pi_0 = c_0 \prod_{i>0} \exp[-\beta U_\mathrm{bend}(\theta_{i0})] \, , 
\end{align}
where $c_0$ is the normalization factor. The trial moves, $K_i$, employed by the Markov  chain with transition matrix $T$ act as follows. One of the $n_b$ dynamic branches is chosen with uniform probability and rotated by a random angle chosen within $-\pi/2$ and $\pi/2$ around a random unit vector centered at the center of the molecule. We consider a Markov matrix $T$ that asymptotically samples $\pi$,  such that $\pi=\pi_T$. Explicit details on the model are given in ~\ref{Sec:ModMol}.

In Fig.~\ref{Fig4} we test the algorithms by sampling the dihedral angle, $\omega_{ij}$, defined as the angle between the plane spanned by the branches $0$ and  $i$ and the plane spanned by the branches $0$ and $j$. Algorithms A and B  reproduce the target distribution. However, the acceptance rates  could be quite small, especially in the case of the 4--branched molecule, even when using the expensive Algorithm B, see  Fig.~\ref{Fig2} ``BM2" and ``BM3",. Fig.~\ref{Fig4} shows how the distributions of $y_n$ are still far from $\pi$ even for the highest value of $n$ considered.

The previous considerations unveil  a limitation of the method:  In systems with many degrees of freedom, prohibitively large values of $n$ may be required to generate acceptable configurations. One should mention that a reduction of the mixing time can be achieved using a prior distribution overlapping with $\pi$ (notice that $n_\mathrm{mix}=0$ if $\pi_0 = \pi$). This could be achieved, for instance, by using learned  priors~\cite{sbailo2021neural} or  an iterative scheme, the latter refining $\pi_0$ on the fly. Notice also that in the example of branched molecules most of the proposed configurations were rejected due to a  random choice of the proposed updates,  $\boldsymbol{K}$. Instead, one could envisage using symplectic or learned~\cite{noe2019boltzmann,wu2021unbiased} transformations to generate updates driving far from equilibrium states into the basins of $\pi$ in fewer updates. Investigations in this direction will be addressed in future efforts.  Finally, notice that a reduction of the mixing time could be achieved by employing the presented algorithm to relax a subset of the system's degrees of freedom as done in the heat--bath approach.

\section{Conclusions}\label{Sec:Conc}

This paper studies the problem of generating  trial configurations in MCMC methods by evolving states  sampled from a prior distribution ($\pi_0$). In particular, we consider the possibility of using a Markov  transition matrix $T$ to evolve a configuration  $y_0$ ($y_0 \sim \pi_0$) into $y_n$  ($y_n\sim \pi_n$),  i.e., $y_n=T^n y_0$, and use $y_n$ as a trial configuration. A limiting condition arises from the necessity of using a value of $n$ large enough to guarantee an overlap between $\pi_n$ and the target distribution ($\pi$). We discussed how this could be problematic when using an arbitrary $T$ and $\pi_0$. Sec.~\ref{Sec:HC} provides an example in which this problem  is sidestepped by using multiple Markov  transition matrices to generate a single configuration.

As done in path sampling  methods, we have defined an extended space comprising all the trajectories generated by  $T$ and derived  two prototypical algorithms ( Algorithm A and B). Intriguingly, in Algorithm A, the quality of the sampling is not only controlled by  the overlap between the trial and target distributions ($\rVert \pi_n -\pi \rVert_\mathrm{TV}$) but also by the  overlap between the distribution of the equilibrium and generated trajectories. Algorithm B addresses the last issue at the price of a higher computational cost  (arising from the necessity of enumerating $2^n$ paths) which heavily limits the length $n$ of the  tMC. We anticipate that it is possible to limit the number of trajectories by modifying the generative method of Sec.~\ref{Sec:Generate_y}. Investigations in this direction will be presented elsewhere.

The proposed methods could also inspire new developments in the field of generative models where, traditionally, neural networks are used to map a prior into a target distribution~\cite{noe2019boltzmann}.

\section*{Acknowledgements}

We thank two anonymous Reviewers and Manuel Ath\`enes for insightful comments and constructive suggestions.  We thank Manuel Ath\`enes  for bringing to our attention relevant literature about path sampling  methods. We thank Alessandro Bevilacqua for early discussions on the sampling of stretched harmonic chains which motivated the developments of the proposed methodology. This paper is dedicated to his memory. Financial support was provided by the Universit\'e Libre de Bruxelles (ULB) {  and an {\em A.R.C. grant of the F\'ed\'eration  Wallonie--Bruxelles}.} Computational resources have been provided by the Consortium
des \'Equipements de Calcul Intensif (CECI), funded by the Fonds de la
Recherche Scientifique de Belgique (F.R.S.--FNRS) under Grant No. 2.5020.11. 

\appendix

\section{Numerical recipes}
 In this section, we provide  the flow charts of the algorithms introduced in this work.  In~\ref{Sec:Pn}, we detail the algorithm used to sample the acceptances $\eta^B_i$'s  employed in Algorithm B. These  acceptances are used to calculate $\oPgen$,  ``B'', in Fig.~\ref{Fig3}.

\subsection{ Generating trial configurations}\label{Sec:App_Generate_y}
Flow chart to generate the path ${\boldsymbol{y}_{0,n}}=(y_0, \, y_1, \, \cdots,\, y_n)$ and compute  $\Pgen(\boldsymbol{y}_{0,n}|\boldsymbol{K})$:
\begin{enumerate}[label=(\roman*)]
    \setlength{\itemsep}{0pt}
    \setlength{\parskip}{0pt}
    \setlength{\parsep}{0pt} 
\item Sample $y_0$ from  the prior distribution $\pi_0$;
\item  Sample a set of  $n$ transformations, $\boldsymbol{K}=(K_1,\cdots,K_n)$;
\item Initialize $\Pgen(\boldsymbol{y}_{0,n}|\boldsymbol{K})$, $\Pgen(\boldsymbol{y}_{0,n}|\boldsymbol{K})=\pi_0(y_0)$;
\item Iterate from $i=1$ to $n$ :
\begin{itemize}
    \setlength{\itemsep}{0pt}
    \setlength{\parskip}{0pt}
    \setlength{\parsep}{0pt} 
	\item Generate  $y_i^{\mathrm{trial}}=K_i y_{i-1}$;
	\item Calculate the acceptance rate for $y^\mathrm{trial}_i$ using $\pi_T$,  $\mathrm{acc}^{(T)}(y_{i-1}\rightarrow y^\mathrm{trial}_i)$;
	\item If accepted, $y_i = y^\mathrm{trial}_i$,  $\Pgen(\boldsymbol{y}_{0,n}|\boldsymbol{K})  \leftarrow \Pgen(\boldsymbol{y}_{0,n}|\boldsymbol{K}) \cdot f_{K_i}(y_{i-1},1)$ (Eq.~\ref{Eq:f});
	\item Else, $y_i=y_{i-1}$ and  $\Pgen(\boldsymbol{y}_{0,n}|\boldsymbol{K})  \leftarrow \Pgen(\boldsymbol{y}_{0,n}|\boldsymbol{K})\cdot f_{K_i}(y_{i-1},0) $;
\end{itemize}
\item Return $\boldsymbol{y}_{0,n}$, $\Pgen(\boldsymbol{y}_{0,n}|\boldsymbol{K})$, $\boldsymbol{K}$.
\end{enumerate}

\subsection{Algorithm A}\label{Sec:Pgen}
Flow chart to calculate the acceptance rate starting from the current configuration $x$ (in the physical space):
\begin{enumerate}[label=(\roman*)]
    \setlength{\itemsep}{0pt}
    \setlength{\parskip}{0pt}
    \setlength{\parsep}{0pt} 
\item Generate the  trial configuration $y$ and the associate path $\boldsymbol{y}_{0,n}$, along with the set of transformations $\boldsymbol{K}$ and the generating probability $\Pgen(\boldsymbol{y}_{0,n}|\boldsymbol{K})$ with the steps described in Sec.~\ref{Sec:App_Generate_y};
\item  Sample $n$ random acceptances, $\eta^A_i$'s, where $\eta^A_i=0$ or 1 with equal probability;
\item  Construct $\boldsymbol{x}_{0,n}$ by setting $x_n=x$ and iterating ${x_i}=K_{i+1}^{-\eta^A_{i+1}}(x_{i+1})$, for $i=n-1,\cdots,0$;
\item  Compute the generating probability $\Pgen(\boldsymbol{x}_{0,n}|\boldsymbol{K})$ using Eq.~\eqref{Eq:Pgen}
\begin{align}
 \Pgen(\boldsymbol{x}_{0,n}|\boldsymbol{K})&=\pi_0(x_0)\prod_{j=1}^n  f_{K_{j}}(x_{j-1},\eta^A_j) \label{Eq:PgenOld}
\end{align}
 with $ f_{K_{j}}$ defined in Eq.~\eqref{Eq:f};
\item  Calculate the acceptance rate with $z_A$ given by Eq.~\eqref{Eq:zA}.
\end{enumerate}

\subsection{Algorithm B }\label{Sec:Pn}
Flow chart to calculate the acceptance rate starting from the current configuration $x$ (in the physical space):
\begin{enumerate}[label=(\roman*)]
    \setlength{\itemsep}{0pt}
    \setlength{\parskip}{0pt}
    \setlength{\parsep}{0pt} 
\item Generate $y$ and the set  $\boldsymbol{K}$ with the steps described  in ~\ref{Sec:App_Generate_y};
\item Compute $P_n(x|\boldsymbol{K})$ and $P_n(y|\boldsymbol{K})$ using  $\boldsymbol{K}$ and the method described below;
\item Calculate the acceptance rate with  $z_B$ given by Eq.~\eqref{Eq:zB}.
\end{enumerate}

Flow chart to calculate $P_n(x_n|\boldsymbol{K})$ (or $P_n(y_n|\boldsymbol{K})$) for a given set of transformations ${\boldsymbol{K}}$
\begin{enumerate}[label=(\roman*)]
    \setlength{\itemsep}{0pt}
    \setlength{\parskip}{0pt}
    \setlength{\parsep}{0pt} 
\item Initialise a $2^n$--dimensional vector ($v_\alpha$, $\alpha=0,\cdots , 2^n-1$) with the list of states at layer 0, $x_0^{(\alpha)}$  (see Fig.~\ref{Fig1}{\bf c}, right), leading to $x_n$ with a combination of displacements  taken from $\boldsymbol{K}$. In particular, if $b_1b_2\cdots b_n$ is the binary  representation of $m=2^n+\alpha$,~\footnote{We  calculate the binary representation of $m$ using the routine available at \url{https://www.geeksforgeeks.org/python-slicing-extract-k-bits-given-position}} we calculate $x_0^{(\alpha)}$ as 
\begin{align}
     v_\alpha =  x_0^{(\alpha)} =(K_1^{-b_1})\cdots(K_n^{-b_n})x_n
\end{align}
\item Initialise $P_\alpha$ and $J_\alpha$ ($\alpha=0,\, \cdots , 2^n-1$) as $P_\alpha=J_\alpha=\pi_0(v_\alpha)$, where $\pi_0$ is the prior distribution;
\item Iterate $P_\alpha$ and $J_\alpha$ $n$ times using the following procedure. At the $i^\mathrm{th}$ iteration, update the components of $P_\alpha$ and $J_\alpha$ with $\alpha = p \cdot 2^{i}$, with $p=0,\, 1, \cdots 2^{n-i-1}$ using Eq.~\eqref{Eq:Bayes}.  In particular, defining $V$ and $\gamma$ as follows 
\begin{align}
V &= P_{p \cdot 2^{i}} f_{K_i}(v_{p \cdot 2^{i}},0)
\nonumber \\
&+P_{p\cdot 2^{i} + 2^{i-1} } f_{K_i}(v_{ p \cdot 2^{i}+ 2^{i-1} } ,1)
\nonumber\\
\gamma &= \frac{ {P_{p\cdot 2^{i} + 2^{i-1} }  f_{K_i}(v_{ p \cdot 2^{i}+ 2^{i-1} } ,1 )} }{V}
\nonumber
\end{align}
then
\begin{align}
J_{p \cdot 2^{i}} & \leftarrow 
\left\{
\begin{array}{l}
J_{p \cdot 2^{i}}  f_{K_i}(v_{p \cdot 2^{i}},0)\;,\\ \quad \text{with } \mathrm{Prob}=1-\gamma \\
J_{p\cdot 2^{i} + 2^{i-1} }   f_{K_i}(v_{ p \cdot 2^{i}+ 2^{i-1} } ,1 )\;,\\ \quad \text{with } \mathrm{Prob}=\gamma 
\end{array}
\right.
\nonumber
\\
 P_{p \cdot 2^{i}} & \leftarrow V
\nonumber
\end{align}
 Note that, by construction, $v_{p \cdot 2^{i}} = K_i v_{p \cdot 2^{i} + 2^{i-1} }$ while, in general, $K_i v_{p \cdot 2^{i}}$ (entering the calculation of $f_{K_i}(v_{p \cdot 2^{i}},0)$) does not belong to the set of $2^n$ states listed by $v_\alpha$ (see Fig.~\ref{Fig1}{\bf c}, right).
\item After the $n^\mathrm{th}$ iteration, the value of $P_n(x_n|\boldsymbol{K})$ and $\Pgen(\boldsymbol{x}_{0,n}|\boldsymbol{K})$ are found as  $P_n(x_n|\boldsymbol{K})=P_0$ and $\Pgen(\boldsymbol{x}_{0,n}|\boldsymbol{K})=J_0$.
\end{enumerate}

\section{Prior and target distributions employed}


Throughout this work,  $f_\mathcal{N}(\textbf{x};\boldsymbol{\mu},\boldsymbol{\Sigma})$ is a Gaussian distribution 
\begin{align}
	f_\mathcal{N}(\textbf{x};\boldsymbol{\mu},\boldsymbol{\Sigma})\equiv \frac{\exp\left[-\frac{1}{2}(\textbf{x}-\boldsymbol{\mu})^\top \boldsymbol{\Sigma}^{-1}(\textbf{x}-\boldsymbol{\mu})\right]
}{(2\pi)^{d/2}|\boldsymbol{\Sigma}|^{1/2}}\end{align}
where $\boldsymbol{\mu}$ is the $d$--dimensional mean vector and $\boldsymbol{\Sigma}$ the covariance matrix.

\subsection{ 2D model (Fig.~\ref{Fig1})}\label{Sec:Mod2D}

The prior distribution is a two--dimensional Gaussian distribution
\begin{align}
\pi_0(\textbf{v})=f_\mathcal{N}(\textbf{v};\boldsymbol{0},\boldsymbol{\Sigma}_0)\, ,
\end{align}
with a covariance matrix given by  $\boldsymbol{\Sigma}_0=\text{diag}(\sigma_1^2,\sigma_2^2)$ where $\sigma_1 = \sigma_2 = 0.6$.

 We use a multimodal target distribution given by the sum of three Gaussian distributions 
\begin{align}
\pi(\textbf{v}) & = \frac{1}{3} \sum_{i=1}^3 f_\mathcal{N}(\textbf{v};\boldsymbol{\mu}_i,\boldsymbol{\Sigma}_i) 
\end{align}
where the inverse of the covariance matrix $ \Sigma_i$ is given by
\begin{align}
  \boldsymbol{\Sigma}^{-1}_i &=\frac{1}{\sigma^2_{i,1}\sigma^2_{i,2}(1-\rho_i^2)} \left(\begin{array}{cc} \sigma_{i,2}^2 & -\rho\sigma_{i,1}\sigma_{i,2} \\-\rho\sigma_{i,1}\sigma_{i,2} &  \sigma_{i,1}^2 \end{array}\right)\, ,
\end{align}
with
\begin{align}
    \boldsymbol{\mu}_1 &= (-2,-2)\;, & \rho_1 &= 0.7\;,\nonumber
\\
    \sigma_{1,1} & = 0.5 \;, &  \sigma_{1,2} &= 0.5 \;,\nonumber
\\
    \boldsymbol{\mu}_2 &= (-1,-2)\;,  & \rho_2 &= 0\;,\nonumber
\\
    \sigma_{2,1} & = 0.3  \;, &  \sigma_{2,2} &= 0.6\;,\nonumber
\\
    \boldsymbol{\mu}_3 &= (-2,2)\;,  & \rho_3 &= 0.3\;,\nonumber
\\
    \sigma_{3,1} & = 0.6 \;, &  \sigma_{3,2} &= 0.3\;.
\nonumber
\end{align}
Prior and target distributions are shown on Fig.~\ref{Fig1}{\bf a}.

\subsection{The harmonic chain system (Sec.~\ref{Sec:HC})}\label{Sec:ModHS}

In the system of Sec.~\ref{Sec:HC}, neighboring monomers interact via a harmonic potential, $V_0$, which reads as follows 
\begin{align}
V_0({\bf r}_i, {\bf r}_{i-1}) &= \frac{ k_B T}{2} |{\bf r}_i - {\bf r}_{i-1}|^2\;.
\end{align}
The prior distributions employed to select ${\bf y}^{(j)}_0$ ($j=1, \cdots , N_T-1$)  has a probability density function given by
\begin{align}
\pi_0({\bf y}^{(j)}_0)=f_{\cal N}({\bf y}^{(j)}_0 ; {\bf r}^\mathrm{t}_{j-1}, {\bf 1} )\;.
\end{align}
The asymptotic state visited by  the tMC with transition matrix $T^{(j)}$, $\pi_{T^{(j)}}$, is taken equal to
\begin{align}
\pi_{T^{(j)}}(y^{(j)}) = \frac{ P_G({\bf y}^{(j)},{\bf r}_{N_T}) \pi_0({\bf y}^{(j)})}{ \int \mathrm{d} {\bf y}^{(j)} P_G({\bf y}^{(j)},{\bf r}_{N_T}) \pi_0({\bf y}^{(j)}) }, 
\end{align}
with 
\begin{align}
P_G({\bf y}^{(j)}_0,{\bf r}_{N_T})=f_{\cal N}({\bf y}^{(j)}_0;{\bf r}_{N_T},  \sqrt{N_T - j}\cdot{\bf 1})\, .
\end{align}

\subsection{Branched molecules (Fig.~\ref{Fig4})}\label{Sec:ModMol}
We consider molecules constituted by $n_b+1$ branches ($n_b=2$ and 3, see Fig.~\ref{Fig4}). $\pi$ reads as follows
\begin{align}
    &\pi(\theta_{01},\cdots, \theta_{0n_b}, \theta_{12},\cdots,\theta_{(n_b-1)n_b}) =\notag\\
    &\frac{1}{{N}} \exp\left[-\beta\sum_{i=0}^{n_b}\sum_{j>i}^{n_b} U_{\text{bend}}(\theta_{ij})\right]\;,
\end{align}
where  $\theta_{ij}$ is the angle between the branches $i$ and $j$, and ${N}$ a normalization constant.
The bending potential $U_{\text{bend}}(\theta_{ij})$ is defined as
\begin{align}
    U_{\text{bend}}(\theta) & = \frac{1}{2}k_{\theta} (\theta-\theta_0)^2\;,
\end{align}
where $k_{\theta}$ and $\theta_0$ are parameters of the system (see below).
We choose the following  probability density function for the  prior distribution
\begin{align}
    \pi_0(\theta_{01},\cdots, \theta_{0n_b}) = \frac{1}{{N}_0} \exp\left[-\beta\sum_{i=1}^{n_b}U_{\text{bend}}(\theta_{0i})\right]\;,
\end{align}
where ${N}_0$ is the normalization factor. 
\\
In Sec.~\ref{sec:limitations}, we consider $2$--methylpropane molecules with  parameters
\begin{align}
    n_b & = 2\;,\\
    {\mathrm{T}} & = 300\ K\;,\\
    \theta_0  & = 112\ (\text{deg})\;,\\
    k_{\theta} / k_B & = 62500\ K\;,
\end{align}
as well as $2,2$--dimethylpropane with  parameters
\begin{align}
    n_b & = 3\;,\\
     {\mathrm{T}} & = 300\ K\;,\\
    \theta_0  & = 109.47\ (\text{deg})\;,\\
    k_{\theta} / k_B & = 62500\ K\;.
\end{align}
\\
The dihedral angle $\omega_{ij}$ (see Fig.~\ref{Fig4}) is the angle between the plane spanned by the branches $0$ and  $i$ and the plane spanned by the branches $0$ and $j$. This angle is obtained from the bending angles $\theta_{0i}$, $\theta_{0j}$ and $\theta_{ij}$ as
\begin{align}
    \cos(\omega_{ij}) & = \frac{\cos(\theta_{ij})-\cos(\theta_{0i})\cos(\theta_{0j})}{\sin(\theta_{0i})\sin(\theta_{0j})}\;.
\end{align}

\section{Asymmetrically distributed displacements $\boldsymbol{K}$}
\label{app:ASdisplacement}
\begin{figure}[t]
     \centering
    \includegraphics[width=0.48\textwidth]{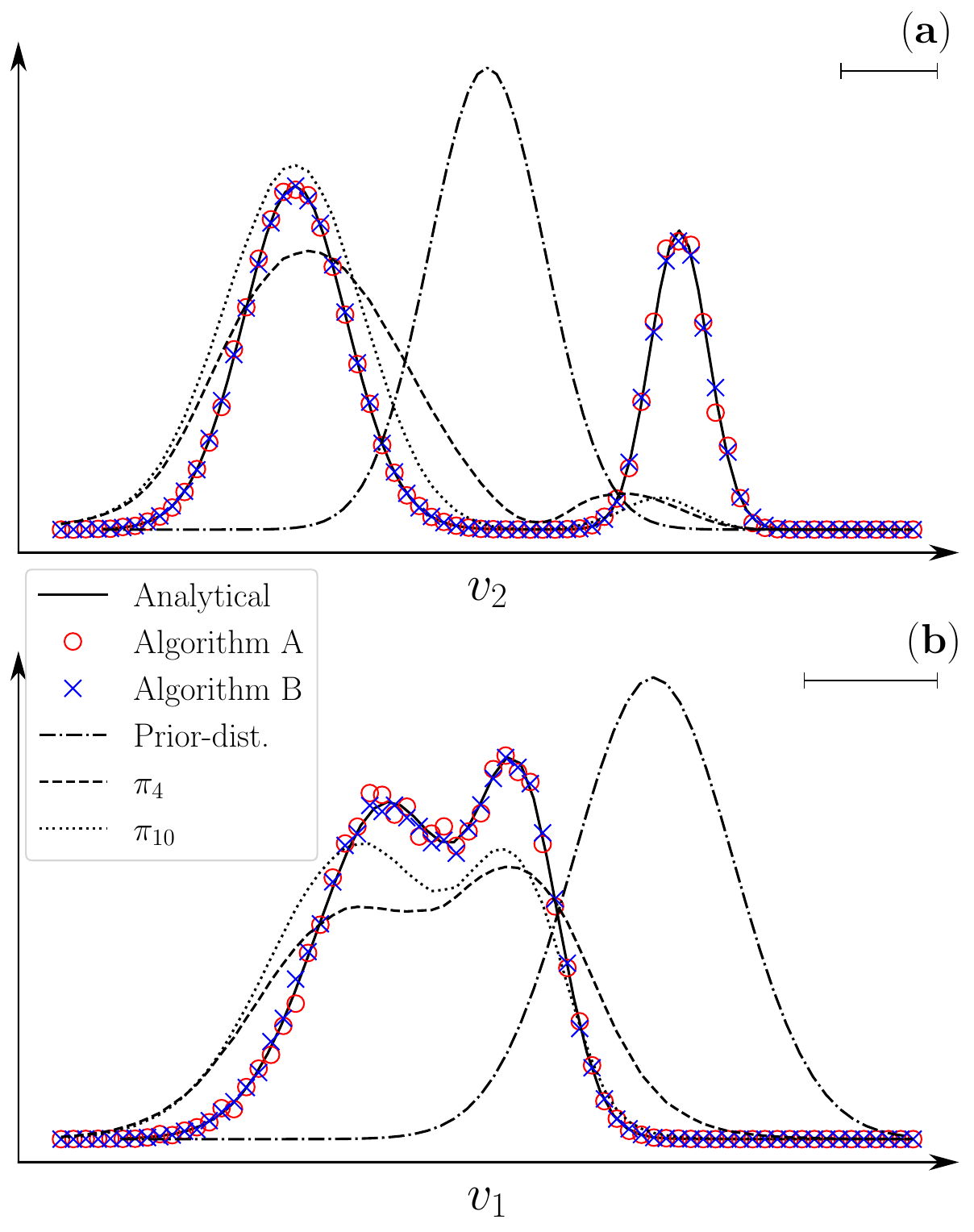}
	\caption{2D model: Comparison of the marginal distributions of $v_1$ ({\bf a}) and $v_2$ ({\bf b}) for Algorithms A and B (symbols) for $n=4$ and asymmetrically distributed displacements $\boldsymbol{K}$, with the analytical target distribution (solid line). Each transformations $K_i$ attempts to displace the current state by  $(\Delta x,\Delta y) \in [-3,1]\times[-3,1]$. The dashed (dotted) lines show the distributions of the  trial configurations for $n=4$ ($n=10$),  The dash--dotted lines depict the prior distribution.  The scale bar represents the unit length. The results have been obtained with $4\cdot10^6$ iterations.}
        \label{FigAS}
\end{figure}

The condition $\mu(\boldsymbol{K})=\mu(\boldsymbol{K}^{-1})$, which is  usually required to enforce the detailed balance condition in path sampling methods, is not required  by Algorithms A and B. To support this statement,  in Fig.~\ref{FigAS} we show that an asymmetric distribution of displacements does not bias the sampling of the  2D model defined in Fig.~\ref{Fig1}{\bf a}.

%

\section{Truncated Markov  chain sampling with $\pi \neq \pi_T$}
\label{app:AlgosABprime}

\begin{figure}[t]
     \centering
    \includegraphics[width=0.48\textwidth]{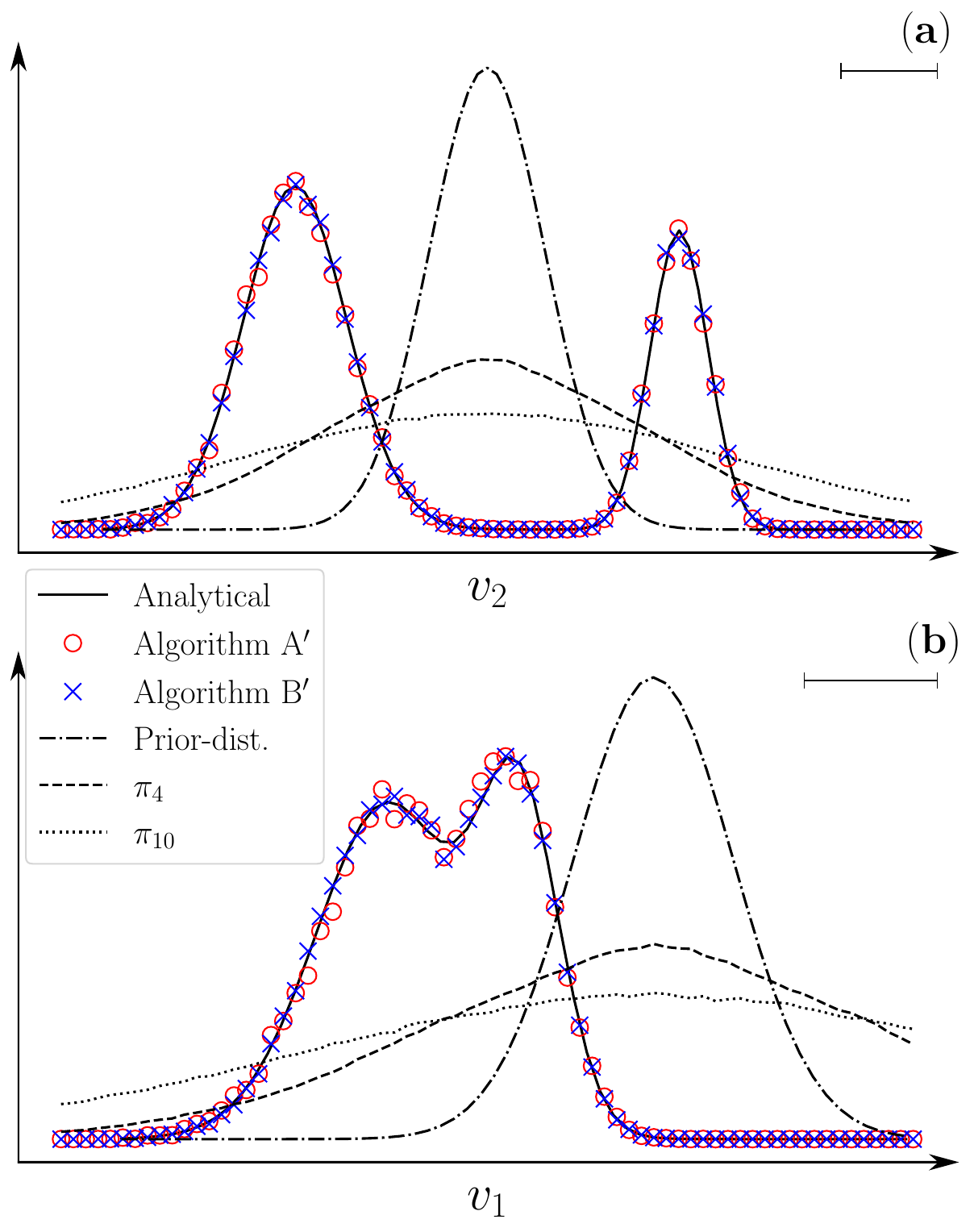}
	\caption{2D model: Comparison of the marginal distributions of $v_1$ ({\bf a}) and $v_2$ ({\bf b}) for Algorithm A' and B', $n=4$, (symbols) with the analytical target distribution (solid line). The dashed (dotted) lines show the distributions of the  trial configurations for $n=4$ ($n=10$). The dash--dotted lines depict the prior distribution.  The results have been obtained with $2\cdot10^6$ iterations.}
        \label{FigSI1}
\end{figure}

\begin{figure}[t]
     \centering
    \includegraphics[width=0.48\textwidth]{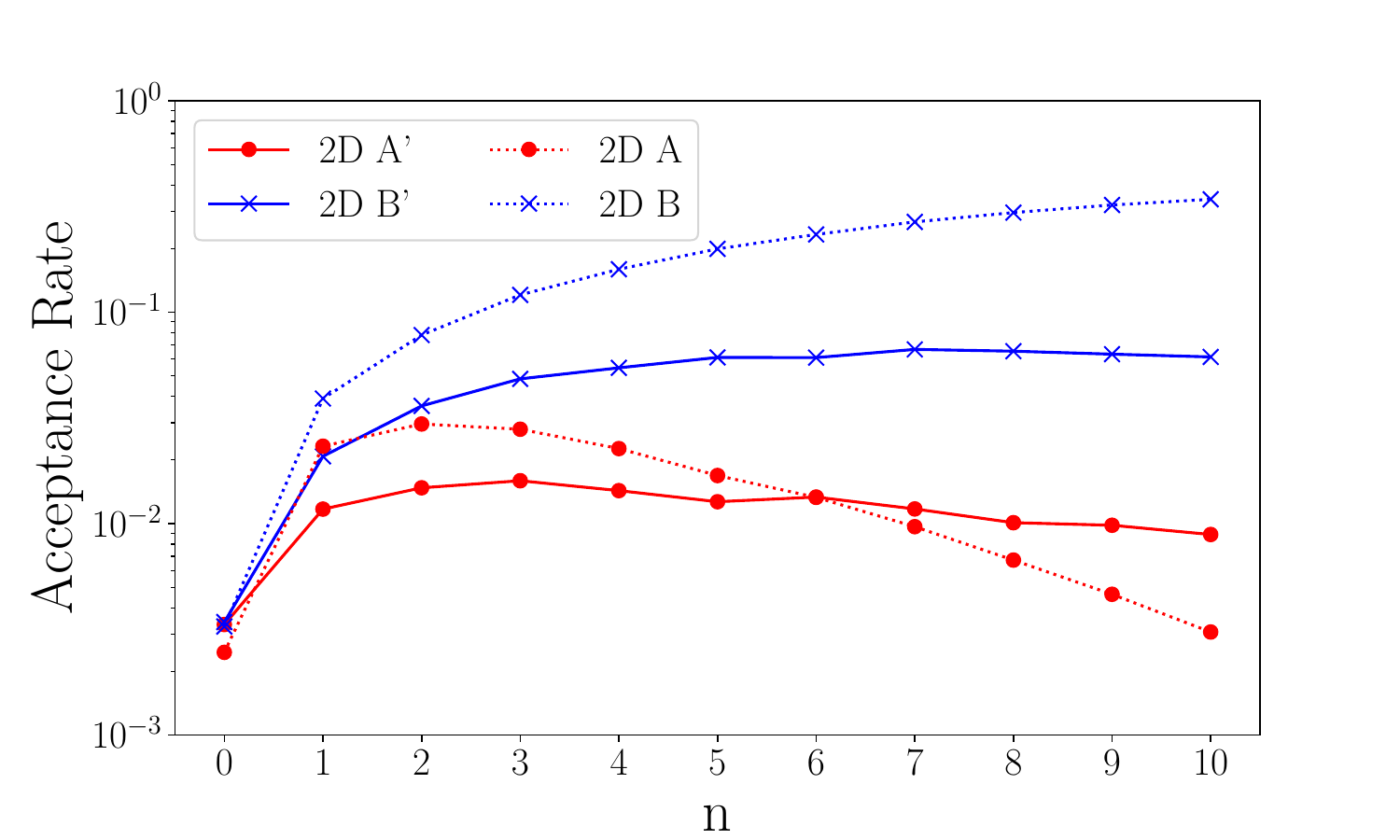}
	\caption{Comparison of the acceptance rates of the Algorithms A' (circles, solid line) and B'(crosses, solid line) with the Algorithms A (circles, dashed line) and B (crosses, dashed line) as function of the number  of $n$ in the   tMC for the two--dimensional system (2D).
}
        \label{FigSI2}
\end{figure}


We consider  a special case of Algorithms A and B where the  trial configurations,  $y_n$, are generated using acceptances $\eta_i$'s uniformly distributed, $\mathrm{Prob}(\eta_i=1)=\mathrm{Prob}(\eta_i=0)=1/2$. In other terms, proposed updates of $y_i$ with $i=0,\cdots, n-1$ are accepted with probability $1/2$, i.e.  $ f_{K_i}=1/2$ for both $\eta=0,1$. The probability of generating a  path leading to $y_n$,  $\boldsymbol{y}_{0,n}=(y_0, \cdots , y_n )$ is then
\begin{align}
    \Pgen( \boldsymbol{y}_{0,n} | \boldsymbol{K})& = \pi_0(y_0) \frac{1}{2^n}\;. \label{Eq:PgenP}
\end{align}
 In this case, the   tMC  would asymptotically sample the constant distribution, $\pi_T\sim 1$. Contrary to  the choice made in the example of Figs.~\ref{Fig1} and~\ref{Fig3} (namely $\pi=\pi_T$), in this section the distribution of the  trial configurations do not attempt to reproduce the target distribution  $\pi$. Below, we  consider Algorithms A' and B', as special cases of Algorithms A and B respectively, where $\Pgen$ is given by Eq.~\eqref{Eq:PgenP}. Using the 2D  model of Fig.~\ref{Fig1}, we show that the Algorithms A' and B' are not biased, supporting that the choice of $\pi_T$ is free.


\subsection{Algorithm A'}\label{Sec:Ap}

Given $x_n$ and $\boldsymbol{K}$, the probability of generating the old configuration  $\boldsymbol{x}_{0,n} = (x_0, \cdots , x_n )$ is given by  
\begin{align}
    \Pgen(\boldsymbol{x}_{0,n} | \boldsymbol{K})& = \pi_0(x_0) \frac{1}{2^n}\;,\label{Eq:PgenxAprime}
\end{align}
using  Eq.~\eqref{Eq:PgenP} and where
\begin{align}
    x_0 =\prod_{j=1}^n K_j^{-\eta^A_j} x_n \;.
\end{align}
The  trial configuration is then accepted with probability $F(z_{A'})$ with 
\begin{align}
    z_{A'}=\frac{ \Pgen(\boldsymbol{x}_{0,n} | \boldsymbol{K}) }{ \Pgen(\boldsymbol{y}_{0,n} | \boldsymbol{K}) } \frac{\pi(y_n) }{ \pi(x_n) } = \frac{\pi_0(x_0)}{\pi_0(y_0)}   \frac{ \pi(y_n) }{ \pi(x_n) }\, ,
\label{Eq:zAprime}
\end{align}
using Eqs.~\eqref{Eq:zA} and~\eqref{Eq:PgenxAprime}.

Fig.~\ref{FigSI1} (red circles) shows that Algorithm A' is not biased even if the distribution of $y_n$ (dashed and dotted lines) does not attempt to reproduce the target distribution (solid line). We observe from Fig.~\ref{FigSI2} ``2D A'' and  ``2D A'', that the acceptance rate is reduced compared to Algorithm A for small values of $n$. This behaviour is explained by  a reduced overlap between the distribution of the  trial configurations (Fig.~\ref{FigSI1} dashed and dotted lines) with $\pi$ as compared to Algorithm A, as shown on Fig.~\ref{Fig1}{\bf b}. For larger values of $n$, Algorithm A' outperforms Algorithm A as in the latter case $P_\mathrm{gen}(\boldsymbol{x}_{0,n}|\boldsymbol{K})/P_\mathrm{gen}(\boldsymbol{y}_{0,n}|\boldsymbol{K})$ is smaller.

\subsection{Algorithm B'}\label{Sec:Bp} 

If   $f_{K_i}=1/2$ for $\eta=0$ and $\eta=1$, then $P_n(y_n|\boldsymbol{K})$ becomes 
\begin{align}
    P_n(y_n|\boldsymbol{K}) = \sum_{i=1}^{2^n}\pi_0(y_{0,i})\frac{1}{2^n}\;,
\end{align}
where $y_{0,i}$ with $i\in(1,2^n)$ are the $2^n$ states obtained from $y_n$
\begin{align}
    y_{0,i} =\prod_{j=1}^n (K_j)^{-\eta^{B,i}_j} y_n \;,
\end{align}
for a set of acceptance $\boldsymbol{\eta}^{B,i}=(\eta_1^{B,i},\cdots,\eta_n^{B,i})$. There are $2^n$ of such sets. Similarly for the old configuration $x$ we obtain
\begin{align}
    P_n(x_n|\boldsymbol{K})& = \sum_{i=1}^{2^n}\pi_0(x_{0,i})\frac{1}{2^n}\;,\\
    x_{0,i} &=\prod_{j=1}^n (K_j)^{-\eta^{B,i}_j} x_n \;.
\end{align}
The  trial configuration is accepted with probability $ \acc^{(P)}=F(z_{B'})$ with  (see Eq.~\eqref{Eq:zB})
\begin{align}
    z_{B'}= \frac{P_n(x_n|\boldsymbol{K})}{P_n(y_n|\boldsymbol{K})}   \frac{ \pi(y_n) }{ \pi(x_n) }=\frac{ \sum_{i=1}^{2^n}\pi_0(x_{0,i}) }{ \sum_{i=1}^{2^n}\pi_0(y_{0,i}) } \frac{\pi(y_n) }{\pi(x_n) } \;.
\label{Eq:zBprime}
\end{align}

 We verify that the conclusions made for Algorithm A' also apply here. Fig.~\ref{FigSI1} (blue crosses) shows that the Algorithm B' is not biased but, as shown in Fig.~\ref{FigSI2} (blue crosses), has a lower acceptance rate  as compared to Algorithm B.

\bibliographystyle{elsarticle-num} 
\bibliography{biblio} 

\end{document}